\documentclass[12pt]{article}

\usepackage[height=8.85in,width=6.45in]{geometry}
\usepackage{fancybox}
\usepackage{amsmath,amssymb}
\usepackage{amsthm}
\usepackage{amscd}
\usepackage[all]{xy}
\usepackage{comment}
\usepackage[table,svgnames,psnames]{xcolor}
\usepackage[colorlinks,citecolor=DarkGreen,linkcolor=FireBrick,urlcolor=FireBrick,linktocpage,unicode,psdextra]{hyperref}

\usepackage{times}
\usepackage{courier}
\usepackage{mathtools}
\numberwithin{equation}{section}

\let\bar\overline

\def\Nequals#1{$\mathcal{N}{=}\,#1$}

\newcommand{\bC}{\mathbb{C}}
\newcommand{\bR}{\mathbb{R}}
\newcommand{\bZ}{\mathbb{Z}}

\newcommand{\SL}{\mathrm{SL}}
\newcommand{\SU}{\mathrm{SU}}
\newcommand{\Sp}{\mathrm{Sp}}
\newcommand{\U}{\mathrm{U}}
\newcommand{\SO}{\mathrm{SO}}

\def\ii{\mathrm{i}}

\newtheorem{thm}{Theorem}[section]
\newtheorem{prop}[thm]{Proposition}
\newtheorem{lem}[thm]{Lemma}

\theoremstyle{definition}

\theoremstyle{remark}

\newcommand{\C}{\mathbb{C}}
\newcommand{\R}{\mathbb{R}}
\newcommand{\Z}{\mathbb{Z}}

\newcommand{\Sing}{\mathcal{S}}

\newcommand{\pt}{\mathrm{pt}}

\begin{document}

\begin{titlepage}

\begin{flushright}
WIS/02/16-FEB-DPPA\\
IPMU-16-0022\\
UT-16-9
\end{flushright}

\vskip 2cm

\begin{center}

{\Large\bfseries S-folds and 4d \Nequals3 superconformal field theories}

\vskip 1cm
Ofer Aharony$^1$ and  Yuji Tachikawa$^{2,3}$ \\
with a mathematical appendix by Kiyonori Gomi$^4$
\vskip 1cm

\begin{tabular}{ll}
1. & Department of Particle Physics and Astrophysics,\\
& Weizmann Institute of Science, Rehovot 7610001, Israel\\
 2. & Department of Physics, Faculty of Science, \\
& University of Tokyo,  Bunkyo-ku, Tokyo 113-0033, Japan\\
3.  & Kavli Institute for the Physics and Mathematics of the Universe, \\
& University of Tokyo,  Kashiwa, Chiba 277-8583, Japan\\
4. & Department of Mathematics, Tokyo Institute of Technology, \\
& 2-12-1 Ookayama, Meguro-ku, Tokyo 152-8551, Japan
\end{tabular}

\vskip 1cm

\end{center}


\noindent
S-folds are generalizations of orientifolds in type IIB string theory, such that the geometric identifications are accompanied by non-trivial S-duality transformations.
They were recently used by Garc\'\i a-Etxebarria and Regalado to provide the first construction of four dimensional ${\cal N}{=}3$ superconformal theories. 
In this note, we classify the different variants of these ${\cal N}{=}3$-preserving S-folds, distinguished by an analog of discrete torsion,
using  both a direct analysis of the different torsion classes and the compactification of the S-folds to three dimensional M-theory backgrounds. 
Upon adding D3-branes, these variants lead to different
classes of ${\cal N}{=}3$ superconformal field theories. We also analyze the holographic
duals of these theories, and in particular clarify the role of discrete gauge and global
symmetries in holography.

In the main part of the paper, certain properties of cohomology groups associated to the S-folds were conjectured and used. 
This arXiv version includes an appendix written by Kiyonori Gomi in 2023 providing the proofs of the required properties using the technique of Borel equivariant cohomology, 
whose brief review is also provided.

\end{titlepage}

\setcounter{tocdepth}{2}
\tableofcontents

\section{Introduction and summary of results}

Field theories with superconformal symmetry are useful laboratories for learning about the behavior of quantum field theories in general, and strongly coupled field theories in particular. This is because the superconformal symmetry allows many computations to be performed in these theories, using methods such as localization, integrability, and the superconformal bootstrap.

The field theories (above two dimensions) about which the most is known are ${\cal N}=4$ superconformal field theories (SCFTs) in four dimensions, that have been called the `harmonic oscillator of quantum field theories'. These theories have an exactly marginal deformation, and it is believed that they are all gauge theories with some gauge group $G$, such that the exactly marginal deformation is the gauge coupling constant. In these theories many observables have already been computed as functions (trivial or non-trivial) of the coupling constant, and there is a hope that they can be completely solved.

Four dimensional theories with ${\cal N}=2$ superconformal symmetry have also been extensively studied. Some of these theories have exactly marginal deformations and corresponding weak coupling limits (at least for some sector of the theory), while others do not. We do not yet have a full classification of ${\cal N}=2$ SCFTs (for rank-1 theories, see \cite{Argyres:2015ffa,Argyres:2015gha,Argyres:2016xua}),
though a large class of theories, called class S, has been constructed following \cite{Gaiotto:2009we}.
Many observables can be computed also in ${\cal N}=2$ SCFTs, at least if they have a weak coupling limit, they are connected by renormalization group flows to theories that have such limits, or they are one of the theories of class S.

Four dimensional ${\cal N}=3$ theories should naively provide an intermediate class of theories, that is more general than ${\cal N}=4$, but such that more computations can be made than in general ${\cal N}=2$ theories. ${\cal N}=3$ SCFTs (that are not also ${\cal N}=4$ SCFTs) have no exactly marginal deformations \cite{Aharony:2015oyb,Cordova:2016xhm} and thus no weak coupling limits that would aid in classifying and performing computations in these theories. Until recently no ${\cal N}=3$ SCFTs were known, but recently a class of such theories was constructed by Garc\'\i a-Etxebarria and Regalado in \cite{Garcia-Etxebarria:2015wns}. Their construction uses a generalization of orientifolds in string theory. 

Orientifold 3-planes (the generalization to other dimensions is straightforward) are defined in type IIB string theory as planes in space-time, such that in the transverse space $y \in \bR^6$ to these orientifolds, there is an identification between the points $y$ and $(-y)$, but with opposite orientations for strings (or, equivalently, with an opposite value for the $B_2$ and $C_2$ 2-form potentials of type IIB). This means that we identify configurations related by a $\bZ_2$ symmetry that involves a spatial reflection in the transverse $SO(6)$, and also a transformation $(-I)$ in the $SL(2,\bZ)$ S-duality group of type IIB string theory. This breaks half of the supersymmetry, preserving a four dimensional ${\cal N}=4$ supersymmetry. In particular, putting $N$ D3-branes on the orientifold (these do not break any extra supersymmetries) gives at low energies four dimensional ${\cal N}=4$ SCFTs.

In \cite{Garcia-Etxebarria:2015wns} this was generalized to identifying configurations related by a $\bZ_k$ symmetry, that acts both by a $(2\pi/k)$ rotation in the three transverse coordinates in $\bC^3 = \bR^6$, and by an element of $SL(2,\bZ)$ whose $k$'th power is the identity. We will call the fixed planes of such transformations S-folds\footnote{%
This term, first coined in \cite{Hull:2004in}, generalizes the term T-folds that is used to describe identifications by elements of the T-duality group. S-folds of string theory involving S-duality twists together with shifts along a circle were studied, for instance, in \cite{Dabholkar:2002sy,Hull:2003kr,Hull:2006tp,CatalOzer:2006mn,ReidEdwards:2006vu}, and similar S-folds were studied in the 4d \Nequals4 SYM theory in \cite{Ganor:2008hd,Ganor:2010md,Ganor:2012mu}.}; for $k=2$ they are the same as the usual orientifolds. Viewing $SL(2,\bZ)$ as the modular group of a torus, the $\bZ_k$ S-duality transformation may be viewed as a rotation of the torus by an angle $(2\pi / k)$; such a rotation maps the torus to itself if and only if $k=3,4,6$ and its modular parameter is $\tau = e^{2 i \pi / k}$, so S-folds of this type exist only for these values of $k$ and $\tau$. It is natural to define such an identification using F-theory \cite{Vafa:1996xn}, in which the $SL(2,\bZ)$ S-duality group is described as adding an extra zero-size torus whose modular parameter is the coupling constant $\tau$ of type IIB string theory; in this language the S-folds of \cite{Garcia-Etxebarria:2015wns} are the same as F-theory on $(\bC^3 \times T^2) / \bZ_k$. This specific $\bZ_k$ identification preserves a four dimensional ${\cal N}=3$ supersymmetry. So, putting $N$ D3-branes on the S-fold gives at low energies theories with ${\cal N}=3$ superconformal symmetry. Note that D3-branes sitting on the S-fold are invariant under all the transformations discussed above, and, in particular, the D3-brane charge of the S-fold is well-defined.

In this note we analyze further the theories constructed in \cite{Garcia-Etxebarria:2015wns}. We focus on asking what are the extra parameters associated with S-folds, analogous to discrete torsion for orbifolds and orientifolds (namely, to non-trivial configurations of the various $p$-forms of type IIB string theory in the presence of the S-fold). S-folds with different parameters can carry different D3-brane charges. Upon putting $N$ D3-branes on the S-fold, these discrete parameters label different ${\cal N}=3$ SCFTs. Our main result is a classification of these extra parameters (following a preliminary discussion in \cite{Garcia-Etxebarria:2015wns}); we show that there are two variants of S-folds with $k=3,4$ and just a single variant with $k=6$. Each variant leads to different ${\cal N}=3$ SCFTs, with different central charges and chiral operators. 

In some cases discrete global symmetries play an interesting role. In a specific type of orientifold (the $O3^-$ plane), the theory of $N$ D3-branes on the orientifold is an $\mathrm{O}(2N)$ gauge theory, and this may be viewed as an $\mathrm{SO}(2N)$ gauge theory in which a discrete global $\bZ_2$ symmetry is gauged. Both the $\mathrm{SO}(2N)$ and the $\mathrm{O}(2N)$ theories exist as ${\cal N}=4$ SCFTs, and a few of their properties are different. We will see that a similar phenomenon happens also for D3-branes on S-folds, and discuss its realization in the AdS/CFT dual of these theories, in terms of discrete gauge symmetries in the bulk. This dual is obviously given by F-theory on $AdS_5 \times (S^5 \times T^2) / \bZ_k$.

We begin in section \ref{sec2} with a naive analysis of the possible discrete parameters, by considering the possible discrete identifications on the moduli space of $N$ D3-branes, and the effects of discrete symmetries. In section \ref{sec3} we discuss the holographic duals of the ${\cal N}=3$ SCFTs, and the realization of the discrete symmetries there. We show that holography suggests that only some of the possibilities found in the naive analysis are consistent. The compactification of F-theory on $(\bC^3 \times T^2) / \bZ_k$ on a circle gives M-theory on $(\bC^3 \times T^2) / \bZ_k$, with each S-fold splitting into several $\bC^4/\bZ_l$ singularities in M-theory. In section \ref{sec4} we show that the consistency of this reduction implies that indeed only the possibilities found in section \ref{sec3} are consistent. In section \ref{sec5} we discuss the fact that for specific values of $k$ and $N$ some variants of the ${\cal N}=3$ theories have enhanced ${\cal N}=4$ supersymmetry, with gauge groups $SU(3)$, $SO(5)$ and $G_2$, and check the consistency of this. This provides new brane constructions and new (strongly coupled) AdS duals for these specific ${\cal N}=4$ SCFTs.

In this paper we discuss only the S-folds which give rise to four dimensional ${\cal N}=3$ supersymmetric theories. It would be interesting to study S-folds that preserve different amounts of supersymmetry, and that have different dimensions. A particularly interesting case, which should have many similarities to our discussion, is S-folds preserving four dimensional ${\cal N}=2$ supersymmetry. We leave the study of these theories to the future.

\paragraph{Note added in v3:}
In the M-theoretic analysis in section~\ref{sec4}, certain properties of cohomology groups associated to the S-folds, namely \eqref{aaaa} and \eqref{bbbb}, are conjectured and used extensively.
In 2023, Kiyonori Gomi kindly proved  these properties,
and prepared a set of notes explaining the details of the proofs,
including a brief review of  the technique of Borel equivariant cohomology,
which is heavily used in it.
The notes are now included as the appendix \ref{appendix} of this arXiv version.

\section{Preliminary analysis}
\label{sec2}

In this paper we study the S-folds introduced by
Garc\'\i a-Etxebarria and Regalado in \cite{Garcia-Etxebarria:2015wns}, which are equivalent to F-theory on $(\bC^3\times T^2)/\bZ_k$ for $k=3,4,6$, generalizing the standard orientifold 3-planes that arise for $k=2$. 
Just as the orientifold 3-planes have four variants, $O3^-$, $\widetilde{O3}{}^-$, $O3^+$ and $\widetilde{O3}{}^+$, that differ by discrete fluxes, we expect that these new S-folds could also have a few variants. 
In this section we perform a preliminary analysis of the possible variants,
from the viewpoint of the moduli space of the \Nequals3 superconformal field theories realized on  $N$ D3-branes probing these S-folds. In the following sections we will discuss additional constraints on the possible variants.

\subsection{Three general properties of ${\cal N}=3$ SCFTs}

We start from the following three properties of theories of this type, that are generally believed to hold:
\begin{itemize}
\item The conformal anomalies (central charges) $a$ and $c$ are equal in any \Nequals3 SCFT \cite{Aharony:2015oyb}: 
\begin{equation}
a=c.\label{a=c}
\end{equation}
The value of $a=c$ determines also all anomalies of the $SU(3)_R\times U(1)_R$ symmetries in these theories.
\item The geometry of the gravitational dual of $N$ D3-branes sitting on a singularity involving a $\bZ_k$ identification of their transverse space, and carrying $\epsilon$ units of D3-brane charge, is AdS$_5\times S^5/\bZ_k$, with $(N+\epsilon)$ units of 5-form flux. This implies that the large $N$ central charge of the corresponding CFTs is 
\begin{equation}
a\sim c \sim k(N+\epsilon)^2/4 + O(N^0).\label{largeNac}
\end{equation}
\item In any \Nequals2 theory, the Coulomb branch operators (chiral operators whose expectation values label the Coulomb branch of the moduli space, namely the space of vacuum expectation values of the scalars in the vector multiplets) form a ring generated by $n$ operators, whose expectation values are all independent without any relation.  Furthermore, up to a caveat mentioned below, there is a relation
\begin{equation}
2a-c =\sum_{i=1}^n (2\Delta_i-1)/4, \label{2a-c}
\end{equation}
where $\Delta_i$ are the scaling dimensions of the generators of the Coulomb branch operators.
\end{itemize}

The property \eqref{largeNac} can be derived easily following the analysis of \cite{Fayyazuddin:1998fb,Aharony:1998xz,Aharony:2007dj}. 
Essentially, the curvature of the AdS space is supported by the total energy of the five-form field, and the volume of the $\bZ_k$ quotient is $1/k$ of the volume before the quotient, changing the value of Newton's constant on $AdS_5$ by a factor of $k$. 

The equation \eqref{2a-c}, originally conjectured in \cite{Argyres:2007tq}, was given a derivation that applies to a large subclass of \Nequals2 theories  in \cite{Shapere:2008zf} (though it is not clear that this subclass includes the theories we discuss here). More precisely, this relation applies only to gauge theories whose gauge group has no disconnected parts.
To illustrate this, consider the \Nequals4 super Yang-Mills theories with gauge groups $\U(1)$ and $\mathrm{O}(2)$.
They have the same central charges $2a-c=1/4$, coming from a single free vector multiplet, but the gauge-invariant Coulomb branch operator has dimension $1$ for the former and $2$ for the latter,
so that property 3 only holds for the former. This is because the scalars in the vector multiplet change sign under the disconnected component of the $\mathrm{O}(2)$ gauge group. The $\mathrm{O}(2)$ theory is obtained by gauging a $\bZ_2$ global symmetry of the $\U(1) = \mathrm{SO}(2)$ theory, where the $\bZ_2$ symmetry is generated by the quotient  $\mathrm{O}(2)/\mathrm{SO}(2)$; this operation does not change the central charges.
This means that the $\mathrm{O}(2)$ theory includes a $\bZ_2$ gauge theory, implying that it has a non-trivial $\bZ_2$ 2-form global symmetry in the sense of \cite{Gaiotto:2014kfa}, as we further discuss below.
We assume in the following that the relation \eqref{2a-c} holds whenever the theory does not have any non-trivial 2-form global symmetry, even if the theory does not have a gauge theory description.

\subsection{D3-branes on S-folds and their discrete symmetries}

Let us now see what the three properties recalled above tell us about the properties of S-folds. 
Let us probe the $\bZ_k$ S-fold by $N$ D3-branes. The moduli space of ${\cal N}=3$ theories is described by the expectation values of scalars in ${\cal N}=3$ vector multiplets, which are identical to ${\cal N}=4$ vector multiplets; viewing an ${\cal N}=3$ theory as an ${\cal N}=2$ theory, these decompose into a vector multiplet and a hypermultiplet. 
Let us choose a specific  \Nequals2 subgroup of the \Nequals3 symmetry, and consider the component of the moduli space which is a Coulomb branch from the point of view of this \Nequals2 subgroup. This implies that all the D3-branes
lie on a particular $\bC/\bZ_k$ within $\bC^3/\bZ_k$.
Denote by $z_i$ $(i=1,\ldots,N)$ the positions on $\bC$ of these D3-branes.
Since the D3-branes are identical objects, the identifications that are imposed on the moduli space by the geometry are 
\begin{align}
z_i &\leftrightarrow z_j, & \text{all other $z_a$ fixed}, \nonumber\\
z_i &\mapsto \gamma z_i, & \text{all other $z_j$ fixed}, \label{oident}
\end{align}
where $\gamma\equiv \exp(2\pi \ii/k)$. For the purpose of the moduli space identifications we can ignore the additional S-duality transformation.

From these identifications we can see that the gauge-invariant Coulomb branch operators are the symmetric polynomials of $z_i{}^k$, and their generators are ($\sum_{i=1}^N z_i{}^{j k}$) for $j=1,\cdots,N$, with dimensions 
\begin{equation} \label{ocoulomb}
k, 2k, \cdots, Nk. 
\end{equation}
From \eqref{a=c} and \eqref{2a-c}, this naively implies that \begin{equation}
4a=4c=kN^2 + (k-1) N.
\end{equation} Applying \eqref{largeNac},  we find that the D3-brane charge of the $\bZ_k$ S-fold is $\epsilon=(k-1)/(2k)$. 

As discussed above, we know that even for $k=2$ this is not the whole story, due to the possibility of discrete torsion and discrete symmetries.
Let us review in detail this $k=2$ case, in a language that will be useful for our later analysis.
Among the known O3-planes, the above analysis only applies to $\widetilde{O3}{}^-$, $O3^+$ and $\widetilde{O3}{}^+$, for which $\epsilon=(k-1)/(2k)=+1/4$ is the correct value.
When the orientifold is $O3^-$, the $\mathrm{O}(2N)$ gauge group arising on $N$ D3-branes can be viewed as a $\bZ_2$ gauging of the $\mathrm{SO}(2N)$ gauge theory, and thus, to compute $a$ and $c$, we need to use the $\mathrm{SO}(2N)$ theory instead of the $\mathrm{O}(2N)$ theory. In the $\mathrm{SO}(2N)$ theory 
the residual gauge symmetries on the Coulomb branch, acting on the eigenvalues of a matrix in the adjoint representation of $\mathrm{SO}(2N)$, are generated by 
\begin{align}
z_i &\leftrightarrow z_j, & \text{all other $z_a$ fixed},\nonumber \\
(z_i,z_j) &\mapsto (-z_i,-z_j), & \text{all other $z_a$ fixed}. \label{soident}
\end{align}
The dimensions of the independent gauge-invariant operators are then \begin{equation}
2, 4, \cdots, 2(N-1); N,
\end{equation} where the last invariant is the Coulomb branch expression for the Pfaffian of an adjoint matrix of $SO(2N)$, 
$z_1z_2\ldots z_N$. 
From this list we can compute $a$ and $c$ again using \eqref{a=c} and \eqref{2a-c} and find \begin{equation}
4a=4c=2N^2 -N,
\end{equation} from which we find $\epsilon=-1/4$ from \eqref{largeNac}.
This is indeed the correct value for $O3^-$.

The $\mathrm{SO}(2N)$ theory has a discrete $\bZ_2$ global symmetry, corresponding to gauge transformations by elements of $\mathrm{O}(2N)$ which are not elements of $\mathrm{SO}(2N)$. Note that in the $\mathrm{SO}(2N)$ theory these are global symmetries rather than gauge symmetries. On the moduli space one of these transformations acts as $z_1 \to -z_1$, and with this extra identification the group generated by \eqref{soident} becomes the same as the group \eqref{oident} acting on the D3-branes. In particular, this identification projects out the Pfaffian operator of dimension $N$, such that after it we obtain the Coulomb branch operators with dimensions \eqref{ocoulomb}. So, for this specific orientifold plane, the theory on the D3-branes is described by a $\bZ_2$ quotient of some `parent' theory which has a different group of identifications \eqref{soident}, and correspondingly different Coulomb branch operators. Only after gauging a $\bZ_2$ global symmetry of this `parent' theory do we get the theory of the D3-branes on the $O3^-$ orientifold.
We stress here that this gauging of $\bZ_2$ symmetry does not change the central charges.

It is natural to ask how we could know directly from the theory of the D3-branes on the $O3^-$ plane that its central charges are not the same as the naive ones, because it arises as a $\bZ_2$ gauging of some other theory. As discussed in \cite{Banks:2010zn,Gaiotto:2014kfa}, when we have a discrete $\bZ_p$ global symmetry, we have local operators that transform under this symmetry, as well as 3-plane operators that describe domain walls separating vacua that differ by a $\bZ_p$ transformation. When we gauge the $\bZ_p$ global symmetry, these local and 3-plane operators disappear from the spectrum. Instead we obtain new 2-plane operators (that may be viewed as worldvolumes of strings), characterized by having a $\bZ_p$ gauge transformation when we go around them. These 2-plane operators are charged under a 2-form $\bZ_p$ global symmetry in the language of \cite{Gaiotto:2014kfa}. So whenever we have a theory with a 2-form $\bZ_p$ global symmetry, it is natural to expect that it arises by gauging a $\bZ_p$ global symmetry of some `parent' theory. And indeed, the analysis of \cite{Gaiotto:2014kfa} implies that this `parent' theory can be obtained by gauging the 2-form $\bZ_p$ global symmetry. Thus, whenever we have a theory with a $\bZ_p$ 2-form global symmetry, we expect that its central charges would not be given by \eqref{2a-c}, but rather by those of its `parent' theory\footnote{Gauging a discrete $\bZ_p$ global symmetry does not change the dynamics on $\bR^4$, but it does change the spectrum of local and non-local operators.}.

Suggested by this analysis for $k=2$, we expect that also for $k=3,4,6$, different versions of S-folds will be characterized by different 2-form $\bZ_p \subset \bZ_k$ global symmetries for the corresponding theories on the D3-branes, that will imply that these theories arise from `parent' theories with $\bZ_p$ global symmetries. The analysis above suggests that the identifications on the moduli spaces of these `parent' theories should be subgroups of the group generated by \eqref{oident}, such that the ring of invariants is polynomial without any relation, and such that adding an extra $\bZ_p$ generator produces the group \eqref{oident}.
Luckily such groups are already classified and are known as Shephard-Todd complex reflection groups; the fact that the ring of invariants is polynomial without any relation if and only if the group acting on $\bC^N$ is a complex reflection group is known as the theorem of Chevalley, Shephard and Todd.\footnote{See \cite{Cecotti:2015hca} for a recent use of Shephard-Todd complex reflection groups in four-dimensional \Nequals2  theories.} For more on complex reflection groups, see \cite{LehrerTaylor}.

In our case, the available groups are known as $G(k,p,N)$, and are generated by the following elements:
\begin{align}
z_i &\leftrightarrow z_j, & \text{all other $z_a$ fixed},\nonumber\\
(z_i,z_j) &\mapsto (\gamma z_i,\gamma^{-1} z_j), & \text{all other $z_a$ fixed},\nonumber\\
z_i & \mapsto \gamma^p z_i, & \text{all other $z_j$ fixed},\label{newident}
\end{align} 
where $p$ is a divisor of $k$. By adding another $\bZ_p$ generator acting as $z_i \mapsto \gamma z_i$, all these groups become the groups \eqref{oident}.
Denoting $\ell=k/p$, we see that if there is a `parent' theory with the identifications \eqref{newident} on its moduli space, then its gauge-invariant Coulomb branch operators would be generated by the symmetric polynomials of $z_i^k$ and by $(z_1z_2\cdots z_N)^{\ell}$, and therefore the dimensions of the Coulomb branch generators would be given by \begin{equation}
k, 2k, \ldots, (N-1)k; N\ell. \label{dims}
\end{equation}
From \eqref{a=c} and \eqref{2a-c} we then find that the central charges of the `parent' theory, and also of its $\bZ_p$ quotient that would describe the theory of $N$ D3-branes on the corresponding S-fold, are given by
\begin{equation}
4a=4c=kN^2+(2\ell-k-1)N.\label{ackl}
\end{equation} 
Note that even though the $\bZ_p$-gaugings of the theories with different values of $\ell$ all have the same moduli space, they are distinct theories with different central charges. When $p$ is not prime, one can also gauge subgroups of $\bZ_p$, giving rise to additional theories, which again are not equivalent to the theories for other values of $\ell$ (even though they may have the same moduli space).

Therefore, it seems  at this stage, that each $\bZ_k$ S-fold with $k=2,3,4,6$ can have variants labeled by $\ell$ which is an integer dividing $k$. 
The  D3-brane charges $\epsilon_{k,\ell}$ of these S-folds can be easily found using \eqref{largeNac} and \eqref{ackl}: \begin{equation}
\begin{array}{c||c|c|c|c|ccc}
&\ell=1 &\ell=2 & \ell=3 & \ell=4 & \ell=6 \\
\hline
\hline
k=2 & \cellcolor{black!10} -1/4 & \cellcolor{black!10}+1/4 &&\\
k=3 & \cellcolor{black!10}-1/3 & & \cellcolor{black!10}+1/3 &&\\
k=4 & \cellcolor{black!10}-3/8 & -1/8 & & \cellcolor{black!10}+3/8& \\
k=6 & \cellcolor{black!10}-5/12 & -1/4 & -1/12 & &+5/12
\end{array}\label{variants}
\end{equation}

Note that just from the analysis above it is far from clear that S-folds and ${\cal N}=3$ theories corresponding to such identifications actually exist. However, we expect any S-fold to fall into one of these categories.
In the next two sections, we will see that the S-fold variants that really exist are only those shaded in the table \eqref{variants} above, first by carefully studying the holographic dual and then by comparing with M-theory. Note also that there could be more than one theory with the same identifications. For $k=2$ there are three orientifold-types with the same $\ell=2$ identifications, though they all give rise to the same theory on D3-branes because they are all related by S-duality. We do not know the corresponding situation in our case. We also cannot rule out the existence of ${\cal N}=3$ theories having Coulomb branch operators \eqref{dims} also for the values of $\ell$ and $k$ that do not come from S-folds. It would be interesting to shed further light on these questions, perhaps by a conformal bootstrap analysis of ${\cal N}=3$ SCFTs, or by analyzing the two dimensional chiral rings of the corresponding theories \cite{Beem:2013sza,Nishinaka:2016hbw}.

\section{Holographic dual}
\label{sec3}
\subsection{AdS duals of CFTs with discrete symmetries}\label{sec:discrete}

Theories of quantum gravity are not expected to have any global symmetries (see, for instance, \cite{Banks:2010zn}). There are very strong arguments that this is the case for continuous symmetries, and it is believed to be true also for discrete symmetries. In the AdS/CFT correspondence, this implies that any global symmetry in the CFT should come from a gauge symmetry in anti-de Sitter (AdS) space. 

For continuous global symmetries it is known that indeed they come from gauge fields in the bulk, with a boundary condition that sets their field strength to zero on the boundary. In some cases (like AdS$_4$ \cite{Witten:2003ya}) there is also another consistent boundary condition in which the field strength does not go to zero on the boundary, and its boundary value becomes a gauge field in the CFT. Changing the boundary condition may be interpreted as gauging the global symmetry in the CFT (in cases where this gauging still gives a CFT). 

In this section we discuss the analogous statements for discrete symmetries. We focus on the $AdS_5 \leftrightarrow$ CFT$_4$ case, but the generalization to other dimensions is straightforward. The discussion here is a special case of the discussion in appendix B of \cite{Gaiotto:2014kfa}, generalized to five dimensions, but as far as we know its implications for the AdS/CFT correspondence were not explicitly written down before (see also \cite{HarlowOoguri}).

Consider a $\bZ_k$ gauge symmetry on $AdS_5$ (thought of as a sector of a full theory of quantum gravity on $AdS_5$, that is dual to a 4d CFT). A universal way to describe such a gauge symmetry in five space-time dimensions is by a topological theory of a 1-form $A$ and a 3-form $C$, with an action
\begin{equation} \label{topzk}
{\cal L} = {{i k}\over {2\pi}} A \wedge d C.
\end{equation}
$A$ can be thought of as the gauge field for a $\bZ_k$ symmetry; for example, the action above arises from a $U(1)$ gauge
symmetry that is spontaneously broken to $\bZ_k$.
The forms $A$ and $C$ are both gauge fields, whose field strengths $dA$ and $dC$ vanish by the equations of motion. There are gauge transformations that shift the integrals of $A$ and $C$ over closed cycles by one. A gauge symmetry is just a redundancy in our description, but an invariant property of this theory is that it has line and 3-surface operators, given by $e^{i \oint A}$ and $e^{i \oint C}$, with $A$ and $C$ integrated over closed cycles. And, it has a 1-form $\bZ_k$ global symmetry that multiplies the line operators $e^{i \oint A}$ by $e^{2\pi i / k}$, and a similar 3-form global symmetry. 

When we put such a theory on $AdS_5$, we need to choose boundary conditions; the possible boundary conditions for such topological gauge theories were discussed in \cite{Gaiotto:2014kfa}. The variational principle implies that we need to set to zero $A\wedge C$ along the boundary. If we set to zero $C$ along the boundary, then the boundary value of $A$ gives a gauge field in the dual CFT, corresponding to a $\bZ_k$ gauge symmetry in this CFT. With this boundary condition line operators in the bulk are allowed to approach the boundary and to become line operators in the CFT, while 3-surface operators cannot approach the boundary, but can end on the boundary, giving 2-surface operators in the CFT. This is as expected for a $\bZ_k$ gauge symmetry in four space-time dimensions.

On the other hand, if we set $A$ to zero on the boundary, we obtain a $\bZ_k$ global symmetry in the dual CFT. The line operators ending on the boundary now give local operators in the CFT, and the 1-form global symmetry in the bulk becomes a standard $\bZ_k$ global symmetry in the CFT, under which these local operators are charged. The 3-surface operators going to the boundary give 3-surface operators in the CFT, which are domain walls between different vacua related by $\bZ_k$.

Thus, whenever we have a $\bZ_k$ gauge symmetry on $AdS_5$, there are two natural boundary conditions. One of them gives a conformal field theory with a $\bZ_k$ global symmetry, and the other gives a conformal field theory with a $\bZ_k$ gauge symmetry (and a $\bZ_k$ 2-form global symmetry). The second theory is related to the first one by gauging its $\bZ_k$ global symmetry, and similarly the first one arises from the second by gauging its $\bZ_k$ 2-form global symmetry.
When $k$ is not prime, there are also additional possible boundary conditions, corresponding to gauging subgroups of $\bZ_k$.

In the context of our discussion in the previous section, this implies that the `parent' theories and their $\bZ_k$-gaugings should arise from the same holographic dual, just with different boundary conditions for the $\bZ_k$ gauge fields. 
We will see  in Sec.~\ref{sec:how} how the $\bZ_k$ gauge fields of \eqref{topzk} arise  in F-theory on $AdS_5\times (S^5 \times T^2) / \bZ_k$ from integrals of the type IIB five-form field $F_5$ on discrete cycles.
Note that all this applies already to the $AdS_5\times S^5/\bZ_2$ case discussed in \cite{Witten:1998xy}; typically in that case only the option of having a global $\bZ_2$ symmetry, leading to the $\mathrm{SO}(2N)$ gauge theory, is discussed.

Finally, note that a very similar story occurs already in type IIB string theory on $AdS_5\times S^5$, which includes a topological sector in the bulk corresponding to a 1-form $\bZ_N$ gauge symmetry, with an action ${i N\over 2\pi} B_2 \wedge dC_2$ (where $B_2$ and $C_2$ are the 2-form fields of type IIB string theory) \cite{Gross:1998gk,Aharony:1998qu}; this topological theory was discussed in section 6 of \cite{Gaiotto:2014kfa}. In this case both simple boundary conditions give rise to a 1-form $\bZ_N$ global symmetry in the dual CFT, and the resulting theories are $SU(N)$ and $SU(N)/\bZ_N$ gauge theories \cite{Aharony:2013hda,Kapustin:2014gua,Gaiotto:2014kfa}. In this specific case there is also an option of coupling these theories to a continuous $U(1)$ gauge symmetry, leading to a $U(N)$ theory \cite{Maldacena:2001ss,Belov:2004ht}. One can also quantize the topological theory in ways that do not lead to local field theories \cite{Witten:1998wy}.

\subsection{Discrete torsion}

In \cite{Witten:1998xy}, Witten  showed how to characterize the variants of O3-planes by studying the discrete torsion on $S^5/\bZ_2$.
In this section we generalize this to $S^5/\bZ_k$ S-folds, for $k=3,4,6$.

For $k=2$, the discrete torsion of the NSNS and of the RR three-form field strengths takes values in  $H^3(S^5/\bZ_2,\tilde\bZ)$, where the tilde over $\bZ$ means that the coefficient system is multiplied by $(-1)$ when we go around the $\bZ_2$ torsion 3-cycle of $S^5/\bZ_2$.
For $k=3,4,6$, we have a $\bZ_k$ torsion 3-cycle in $S^5/\bZ_k$, and we have 
an action of an element $\rho \in \SL(2,\bZ)$ on the NS-NS and R-R field strengths in $\bZ\oplus\bZ $ 
when we go around the $\bZ_k$ torsion cycle of $S^5/\bZ_k$. We can choose a specific form for this
$\rho$, given, say, by $\rho = \begin{psmallmatrix}
-1&0\\ 0 &-1
\end{psmallmatrix}$ for $k=2$, $\rho = \begin{psmallmatrix}
-1&-1\\ 1 &0
\end{psmallmatrix}$ for $k=3$, $\rho = \begin{psmallmatrix}
0&-1\\ 1 &0
\end{psmallmatrix}$ for $k=4$, and $\rho = \begin{psmallmatrix}
0&-1\\ 1 &1
\end{psmallmatrix}$ for $k=6$. Note that the eigenvalues of these matrices are $\gamma$ and $\gamma^{-1}$.
The discrete torsion is then given by $H^3(S^5/\bZ_k,(\bZ\oplus\bZ )_\rho)$, and its
computation is standard in mathematics\footnote{The method to compute cohomologies with twisted coefficients is explained in e.g.~Hatcher \cite{Hatcher} Chapter 3.H ; the cell decomposition of generalized lens spaces, of which $S^5/\bZ_k$ is one, is given in the same book, Example 2.43.  Section 5.2.1 of Davis-Kirk \cite{DavisKirk} was also quite helpful. }.

In general, $H^*(S^{2n-1}/\bZ_k, A)$ where $A$ is a $\bZ_k$-module is given by the cohomology of the complex \begin{multline}
C^0   \xrightarrow{1-t} C^1  \xrightarrow{1+t+\cdots+t^{k-1}} C^2   
\xrightarrow{1-t}  \cdots 
\xrightarrow{1+t+\cdots+t^{k-1}} C^{2n-2}  \xrightarrow{1-t} C^{2n-1}   
\end{multline} where all $C^i\simeq A$, $t$ is the generator of $\bZ_k$, and the differential $d$ is alternately given by the multiplication by $1-t$ or by $1+t+\cdots+t^{k-1}$.
It is easy to see that $d^2=(1-t)(1+t+\cdots+t^{k-1})=1-t^k=0$.

When $k=2$, $t$ is just the multiplication by $-1$. Then $1+t=0$ and $1-t=2$, from which we conclude $H^3(S^5/\bZ_2,(\bZ\oplus \bZ)_\rho)=\bZ_2\oplus \bZ_2$, reproducing four types of O3-planes. 

When $k=3,4,6$, the action $\rho$ of the generator of $\bZ_k$  on $\bZ\oplus\bZ$ obeys
$1+\rho+\cdots+\rho^{k-1}=0$, and $\det(1-\rho) = 3,2,1$ for $k=3,4,6$, respectively, and therefore \begin{equation}
H^3(S^5/\bZ_k,(\bZ\oplus\bZ )_\rho)= 
\begin{cases}
\bZ_3 & (k=3),\\
\bZ_2 & (k=4),\\
\bZ_1 &(k=6).
\end{cases}\label{discretetorsion}
\end{equation} 
This gives the discrete torsion groups arising from the 3-form fields of type IIB string theory on these S-folds.

For $k=3$, we have three different possibilities, but
two non-trivial elements of $\bZ_3$ are related by conjugation in $\mathrm{SL}(2,\bZ)$, so up to S-duality transformations there are just two types of S-folds with $k=3$ (in the same sense that up to S-duality there are just two types of O3-planes that give different theories for the D3-branes on them).
For $k=4$, the cohomology is $\bZ_2$, and therefore we expect two types of S-folds.
Similarly,  for $k=6$, there is only one type of S-fold.

\subsection{Generalized Pfaffians}\label{sec:Pfaff}

In the holographic duals of the S-folds, the discrete torsion described above corresponds to discrete 3-form fluxes on the $S^3/\bZ_k$ discrete 3-cycle in $S^5/\bZ_k$, and it affects the spectrum of wrapped branes on this 3-cycle. We can use this to match the discussion of the previous subsection to our analysis of Section 2. 

For $k=2$ this was analyzed in \cite{Witten:1998xy}. There is a $\bZ_2$-torsion three-cycle of the form $S^3/\bZ_2$ within $S^5/\bZ_2$,
and only when the discrete torsion is zero, i.e.~when the 3-plane is $O3^-$, we can wrap a D3-brane on this cycle.
By analyzing the properties of this wrapped D3-brane we find that it corresponds to a dimension $N$ operator in the dual CFT,
which can be naturally identified with the Pfaffian operator, that only exists in the $\SO(2N)$ theory but not in $\SO(2N+1)$ or $\Sp(N)$ theories. Note that, according to the discussions above, this operator exists in the `parent' $\SO(2N)$ theory, but not after we gauge the $\bZ_2$ to get the $\mathrm{O}(2N)$ theory. So in the AdS dual it exists when we choose the boundary condition for the $\bZ_2$ gauge theory in the bulk that gives a $\bZ_2$ global symmetry, corresponding to the `parent' theory, but not for the other boundary condition, that corresponds to the theory on the D3-branes.

The obstruction to wrap D3-branes on $S^3/\bZ_2$ can be understood as follows. 
The NSNS 3-form flux $G$ can in general be in a non-trivial cohomological class. 
But when pulled-back to the worldvolume of a single D3-brane, $G$ is the exterior derivative of a gauge-invariant object $B-F$, where $F$ is the gauge field on the D3-brane,
and therefore the cohomology class $[G]$ should be trivial. The argument for the RR flux is the S-dual of this.

For all $k=3$, $4$, $6$, there is a $\bZ_k$-torsion three-cycle in $S^5/\bZ_k$ of the form $S^3/\bZ_k$.
When the discrete torsion is zero, there is no obstruction to wrapping a D3-brane on this cycle.
The scaling dimension of this wrapped D3-brane can be easily found to be $kN/k=N$.
This matches the scaling dimensions we found for the $\ell=1$ variants of Section 2. So we identify
the S-fold with no discrete torsion with the $\ell=1$ case (either the `parent' theory, or its $\bZ_k$ gauging that gives the theory on the D3-branes, depending on the boundary conditions).

For $k=6$, there is nothing more to discuss, since $H^3(S^5/\bZ_k,(\bZ\oplus\bZ )_\rho)$ itself is trivial, so we do not find any variant except $\ell=1$.

For $k=3$, when the discrete torsion in  $H^3(S^5/\bZ_k,(\bZ\oplus\bZ )_\rho)=\bZ_3$ is non-trivial, we cannot wrap a D3-brane on this cycle. So
this should correspond to the $\ell=3$ variant (with no discrete symmetries).

The most subtle is the $k=4$ case, when the discrete torsion is given by an element in $H^3(S^5/\bZ_k,(\bZ\oplus\bZ )_\rho)=\bZ_2$. We claim
that the non-trivial element of this discrete torsion gives the $\ell=4$ variant of Section 2, not the $\ell=2$ variant.

To see this, it is instructive to recall why in the case $k=2$ we can wrap two D3-branes on $S^3/\bZ_2$ even with the discrete torsion.
Note that it is not just that for $N$ D3-branes wrapping on the same cycle, the triviality of $N\cdot [G]$ suffices. For one thing, if we have two D3-branes wrapping on the same locus with at least $\U(1)\times\U(1)$ unbroken, then $B-F_1$ and $B-F_2$ are both gauge-invariant (where $F_{1,2}$ are the $\U(1)$ gauge fields coming from the first and the second Chan-Paton factor), and therefore $[G]$ still needs to be zero. 

To wrap two D3-branes on $S^3/\bZ_2$ with discrete torsion consistently, one needs to require that the Chan-Paton indices 1,2 are interchanged when we go around the $\bZ_2$ cycle. This means that in fact there is a single connected D3-brane of the shape $S^3$, which is wrapped on $S^3/\bZ_2$ using a 2:1 quotient map. In this particular setting, we know how to judge the consistency of wrapping: on $S^3$, $(B-F)$ is a gauge-invariant object, so $[G]$ should vanish there. What needs to be checked is then to pull back the spacetime $G$ using the map
\begin{equation}
	S^3 \to S^3/\bZ_2 \to S^5/\bZ_2,
\end{equation}
where the first is the 2:1 projection and the second is the embedding. The result is zero, from the trivial fact that $H^3(S^3,\bZ) = \bZ$ does not have non-zero two-torsion elements. 
Thus we see that we can wrap two D3-branes in this way.

Now, let us come back to the $k=4$ case, and try to wrap two D3-branes on $S^3/\bZ_4$ with discrete torsion, which should naively be possible since the discrete torsion lies in $\bZ_2$. Again, if the Chan-Paton structure is trivial, we cannot wrap them.  If we try to do the analogue of the $k=2$ case above, we can try wrapping two D3-branes such that the Chan-Paton indices 1 and 2 are exchanged. Just as above, this is equivalent to wrapping one D3-brane on $S^3/\bZ_2$, covering $S^3/\bZ_4$ with a 2:1 quotient map.
To test the consistency of the wrapping, one needs to pull back the spacetime obstruction on $S^5/\bZ_4$ via \begin{equation}
	S^3/\bZ_2 \to S^3/\bZ_4 \to S^5/\bZ_4.
\end{equation}
One finds that the pull-back is non-trivial,
showing that this embedding is inconsistent.\footnote{In general, the pull-back map from $S^n/\bZ_{ab}$ to $S^n/\bZ_{b}$ is a multiplication map by $1+t+t^2+ ... + t^{a-1}$, where $t$ is the generator of the $\bZ_a$ action that divides $S^n/\bZ_b$ to $S^n/\bZ_{ab}$.}
It is clear that we can wrap four D3-branes on the discrete cycle, so this implies that the theory with this discrete torsion should be the $\ell=4$ variant discussed in Section 2, and that we cannot have S-folds with the $\ell=2$ variant.

This does not prove, we admit, that there are no other non-Abelian configurations on two D3-branes on $S^3/\bZ_4$ that still allow the wrapping. 
But we will see that the choice $\ell=4$ for this discrete torsion variant matches with the M-theory computation below.

\subsection{Realization of the $\bZ_k$ gauge theory in F-theory}\label{sec:how}
After these discussions we can understand how the $\bZ_k$ gauge theory of Section 3.1 arises in F-theory on $S^5/\bZ_k$, when the discrete torsion in $H^3(S^5/\bZ_k,(\bZ\oplus\bZ)_\rho)$ is zero.
Recall that $H_n(S^5/\bZ_k,\bZ)$ is $\bZ_k$ for $n=1$ and $3$, and its generator is $S^1/\bZ_k$ and $S^3/\bZ_k$, respectively.

We can wrap a D3-brane on these cycles to obtain a world 3-cycle and a worldline on $AdS_5$.
(Note that the latter is the generalized Pfaffian particle discussed above.)
Clearly, they both carry $\bZ_k$ charges.
We are then naturally led to identify the worldline as coupling to the 1-form $A$ and the world 3-cycle to the 3-form $C$ in \eqref{topzk}.
For this identification to make sense, it should be the case that if we rotate the generalized Pfaffian around an $S^1$ that has a unit linking number with the world 3-cycle of the D3-brane wrapped on $S^1/\bZ_k$, 
there should be a non-trivial $\bZ_k$ holonomy of unit strength. 

This can be seen as follows. 
A D3-brane wrapped on $S^1/\bZ_k$ creates an $F_5$ flux given by the Poincar\'e dual of its worldvolume. 
In this case this is given by $s \otimes c$, where $s$ is the generator of $H^1(S^1,\bZ)$ where this $S^1$ is linked in $AdS_5$ around the worldvolume of the D3-brane, and $c$ is the generator of $H^4(S^5/\bZ_k,\bZ)=\bZ_k$. 

Now, we use the cohomology long exact sequence associated to the short exact sequence $0\to \bZ \to \bR \to \U(1)\to 0$ to conclude that there is a natural isomorphism \begin{equation}
H^3(S^5/\bZ_k,\U(1))\simeq H^4(S^5/\bZ_k,\bZ)=\bZ_k.
\end{equation}
In physics terms, this means that the discrete $\bZ_k$ field strength (which is $\bZ$-valued) with four legs along $S^5/\bZ_k$ can be naturally identified with the discrete $\bZ_k$ holonomy (which is $\U(1)$-valued) with three legs along $S^5/\bZ_k$. 

From this we see that we have a holonomy of the type IIB $C_4$ field given by the element \begin{equation}
s\otimes c \in H^1(S^1,\bZ)\otimes H^3(S^5/\bZ_k,\U(1)),
\end{equation} which can be naturally integrated on the cycle $S^1\times S^3/\bZ_k$ to give $\exp(2\pi i/k)$.
This means that the generalized Pfaffian particle wrapped on $S^3/\bZ_k$, when carried around the $S^1$ linking the worldvolume of the D3-brane wrapped on $S^1/\bZ_k$, experiences this holonomy.
This is indeed the behavior we expect for the objects charged under the $\bZ_k$ 1-form $A$ and the $\bZ_k$ 3-form $C$ in \eqref{topzk}.

\section{Comparison to M-theory}
\label{sec4}
In this section, we consider M-theory configurations on $(\bC^3\times T^2)/\bZ_k$. We know that we obtain such configurations from any of our S-folds on a circle, using the standard relation between type IIB theory on a circle and M-theory on a torus. However, the opposite is not true, since when we translate some configuration of discrete fluxes in M-theory on $(\bC^3\times T^2)/\bZ_k$ to F-theory on a circle, we could also have some non-trivial action of the shift around the F-theory circle, corresponding to a `shift-S-fold'
where there is a rotation on the transverse $\bC^3$ when we go around the compactified $S^1$. 
By analyzing all possible discrete charges in M-theory and translating them to F-theory,
we learn about all possible variants of S-folds.

\subsection{Discrete fluxes in M-theory on $(\bC^3\times T^2)/\bZ_k$}
\label{subsec4.1}

Let us start by analyzing the possible discrete fluxes in M-theory, which come from 4-form fluxes.
For a given $k$, $(\bC^3\times T^2)/\bZ_k$ has several fixed points of the form $\bC^4/\bZ_{\ell_i}$,
each of which has an associated $H^4(S^7/\bZ_{\ell_i},\bZ)=\bZ_{\ell_i}$ in M-theory \cite{Aharony:2008gk}.
From this viewpoint, the possible discrete charges are given by the orbifold actions on the fixed points,
\begin{equation}
\bigoplus H^3(\bC^4/\bZ_{\ell_i},\bZ)=
\begin{cases}
\bZ_2\oplus\bZ_2\oplus\bZ_2\oplus\bZ_2 & (k=2),\\
\bZ_3\oplus\bZ_3\oplus\bZ_3 &(k=3), \\
\bZ_4\oplus\bZ_4\oplus\bZ_2 &(k=4), \\
\bZ_6\oplus\bZ_3\oplus\bZ_2 &(k=6). 
\end{cases}
\label{M}
\end{equation}

We can alternatively measure the same charges by considering $H^4$ of the `asymptotic infinity' of $(\bC^3\times T^2)/\bZ_k$, which has the form $(S^5\times T^2)/\bZ_k$. 
This is a $T^2$ bundle over $S^5/\bZ_k$, and as such, one can apply the Leray-Serre spectral sequence\footnote{%
For an introduction on the Leray-Serre spectral sequence, see e.g.~\cite{HatcherSS}.
In our case, the computation goes as follows.
The second page $E^{p,q}_2$ of the spectral sequence  is given by $E^{p,q}_2=H^p(S^5/\bZ_k, H^q(T^2,\bZ))$, with the differentials $d_2^{p,q}:E^{p,q}_2\to E^{p+2,q-1}_2$. The third page is given by $E^{p,q}_3=\mathrm{Ker}\ d^{p,q}_3/\mathrm{Im}\ d^{p-2,q+1}$, and has the differentials $d_3^{p,q}:E^{p,q}_3\to E^{p+3,q-2}_3$.
The second page and the part relevant for us of the third page are then given by \[
\begin{array}{c|cccccc}
2& \bZ & 0 &\bZ_k & 0 & \bZ_k & \bZ \\
1 & 0 & X & 0 & X & 0& X \\
0 & \bZ & 0 &\bZ_k & 0 & \bZ_k & \bZ \\
\hline 
& 0 & 1 & 2 & 3 & 4 & 5
\end{array},\qquad
\begin{array}{c|cccccc}
2&  & 0 &\bZ_k &  &  &  \\
1 &  &  &  & X & &  \\
0 &  &  & &  & \bZ_k & \bZ \\
\hline 
& 0 & 1 & 2 & 3 & 4 & 5
\end{array}
\] where $X=\bZ_2\oplus \bZ_2, \bZ_3, \bZ_2, \bZ_1$ for $k=2,3,4,6$, respectively.
From this we conclude that in fact $F^{p,q}/F^{p+1,q-1}=E^{p,q}_\infty \simeq E^{p,q}_2$ for $p+q=4$.}, that says that it has the filtration \begin{equation}\label{filtration}
H^4((S^5\times T^2)/\bZ_k,\bZ) = F^{2,2} \supset F^{3,1} \supset F^{4,0},
\end{equation} where \begin{align}
F^{2,2}/F^{3,1}&=H^2(S^5/\bZ_k, H^2(T^2,\bZ))=H^2(S^5/\bZ_k, \bZ),\label{filone}\\
F^{3,1}/F^{4,0}&=H^3(S^5/\bZ_k, H^1(T^2,\bZ))=H^3(S^5/\bZ_k, (\bZ\oplus\bZ)_\rho),\label{filtwo}\\
F^{4,0}&=H^4(S^5/\bZ_k, H^0(T^2,\bZ))=H^4(S^5/\bZ_k, \bZ).\label{filthree}
\end{align} The different subgroups here correspond in a sense to the `number of legs along the base $S^5/\bZ_k$ and the fiber $T^2$'. 
As we are taking the $\bZ$-valued cohomology, we only have a filtration and it is not guaranteed that the group \eqref{filtration} is a direct sum of \eqref{filone}--\eqref{filthree}.

We can easily compute  \eqref{filone}--\eqref{filthree} to obtain
 \begin{equation}
F^{2,2}/F^{3,1}=\bZ_k,\quad
F^{3,1}/F^{4,0}=
\begin{cases}
\bZ_2\oplus \bZ_2 & (k=2)\\
\bZ_3 & (k=3)\\
\bZ_2 &(k=4)\\
\bZ_1 & (k=6)
\end{cases},\quad
F^{4,0}=\bZ_k.
\end{equation}

From the standard F-theory / M-theory mapping, we see that each piece \eqref{filone}--\eqref{filthree}
has the following effect in the F-theory language:
\begin{itemize}
\item  The piece $F^{2,2}/F^{3,1}$ becomes a component of the metric of F-theory on $S^1$.
Concretely, it specifies the amount of the rotation on the transverse space while we go around $S^1$.
Therefore, the system is a plain $S^1$ compactification if the flux in this piece is zero, whereas it is a shift S-fold if it is non-zero.
\item The piece $F^{3,1}/F^{4,0}$ becomes the discrete RR and NSNS 3-form fluxes around the S-fold.
Indeed, the coefficient system $H^1(T^2,\bZ)$ of the fiber is exactly the one $(\bZ\oplus\bZ)_\rho$ that we discussed in the previous section, and there is a natural isomorphism of $F^{3,1}/F^{4,0}$ with the discrete torsion \eqref{discretetorsion}.
\item The piece $F^{4,0}$ becomes an $F_5$ flux having one leg along the $S^1$. As discussed in Sec.~\ref{sec:how}, we have a natural identification $H^4(S^5/\bZ_k,\bZ)=H^3(S^5/\bZ_k,\U(1))$. 
So, this piece can also be regarded as specifying  a $C_4$ holonomy having one leg along the $S^1$.
\end{itemize}

Note that 
\begin{equation}
(F^{2,2}/F^{3,1})\oplus (F^{3,1}/F^{4,0}) \oplus F^{4,0}
=
\begin{cases}
\bZ_2\oplus\bZ_2\oplus\bZ_2\oplus\bZ_2 & (k=2),\\
\bZ_3\oplus\bZ_3\oplus\bZ_3 &(k=3), \\
\bZ_4\oplus\bZ_4\oplus\bZ_2 &(k=4), \\
\bZ_6\oplus\bZ_3\oplus\bZ_2 &(k=6), 
\end{cases}\label{F}
\end{equation}
and we see that \eqref{F} and \eqref{M} are the same as abstract groups.
What remains is to figure out the precise mapping between the two.
This is a purely geometrical question since it is just the relation of the $H^4$ of $(\bC^3\times T^2)/\bZ_k$ computed at the origin and at the asymptotic infinity. 
As we have not yet done this computation, we will use guesswork\footnote{%
The computation was kindly provided by Kiyonori Gomi in 2023,
whose results are given in Appendix~\ref{appendix} of this arXiv version.
In particular, the crucial relations \eqref{aaaa} and \eqref{bbbb}
below, which were assumed and then used in the rest of this subsection,
were formulated as Proposition~\ref{prop:maps} and proved there.
The main text is kept as originally written in 2016.
}, and then verify that it leads to consistent results for the M2-brane and D3-brane charges of the various singularities.

The $k=2$ case can be worked out using the known properties of the orientifolds. In this case the map \begin{equation}
a: \bigoplus H^3(\bC^4/\bZ_{\ell_i}) =(\bZ_2)^4 \to F^{2,2}/F^{3,1}=\bZ_2
\end{equation} is given by the sum of the four $\bZ_2$'s, and the map \begin{equation}
b: F^{4,0}=\bZ_2 \to \bigoplus H^3(\bC^4/\bZ_{\ell_i})=(\bZ_2)^4 
\end{equation} is the diagonal embedding. 
Note that $a\circ b$ is indeed a zero map.

We assume that a similar relation holds also in the other cases. Namely, we assume that \begin{equation}
a: \bigoplus H^3(\bC^4/\bZ_{\ell_i}) \to F^{2,2}/F^{3,1}=\bZ_k
\label{aaaa}
\end{equation} is the `sum' and that \begin{equation}
 b: F^{4,0}=\bZ_k \to \bigoplus H^3(\bC^4/\bZ_{\ell_i})
 \label{bbbb}
\end{equation} is the `diagonal embedding'. For $k=4$ and $k=6$, we need to use natural maps such as $\bZ_2\to \bZ_6$ and $\bZ_6\to \bZ_2$. We just choose a multiplication by 3 in the former, and the mod-2 map in the latter, and similarly for $k=4$.
We see that the composition $a\circ b$ is indeed a zero map, which gives a small check of our assumptions.

\subsection{M2-brane and D3-brane charges}

 Using the discussions above, let us test our identification by working out the M2-brane and D3-brane charges of our various configurations. 
We use the formula of \cite{Aharony:2009fc} for the charge of $\bC^4/\bZ_k$ orbifolds with discrete torsion,
\begin{equation}
\text{(M2-brane charge)}=-\frac1{24}(k-\frac1k) + \frac{\ell(k-\ell)}{2k}. \label{ahho}
\end{equation}
The cases we use are 
\begin{equation}
\begin{array}{cr|ccccccc}
&&\ell=0 &\ell= 1 &\ell= 2 &\ell= 3 &\ell= 4 &\ell= 5 \\
\hline
k=2: & -\frac1{16}+ & 0 & \frac14 \\[1em]
k=3:& -\frac19 + & 0 & \frac13 & \frac13\\[1em]
k=4: & -\frac5{32} +& 0& \frac38 & \frac12 & \frac38 \\[1em]
k=6: & -\frac{35}{144} + & 0 & \frac5{12} & \frac23 & \frac34 & \frac23 & \frac5{12}
\end{array}.
\end{equation}

A long list of tables analyzing all possible cases follows. 
The end result is that every possible M-theory configuration can be interpreted as a plain $S^1$ compactification or a shift-S-fold on $S^1$ of precisely the S-folds that we discussed in the previous section (the types shaded in \eqref{variants}),
namely the S-folds with \begin{equation}
 (k,\ell)=(2,1) ,  (2,2) ,  (3,1) ,  (3,3) ,  (4,1) ,  (4,4) ,\ \text{and}\  (6,1) .
\end{equation}
In all cases we compute the D3-brane charge using the sum of the M2-brane charges of the orbifold singularities \eqref{ahho} and successfully compare it with our expectation \eqref{variants}.

\subsubsection{$k=2$}
Let us denote the discrete charges of the four fixed points as elements in $\begin{psmallmatrix}
\bZ_2 &\bZ_2\\
\bZ_2 & \bZ_2
\end{psmallmatrix}$. The piece $F_{4,0}=\bZ_2$ is generated by $\begin{psmallmatrix}
0&0\\ 0 &0
\end{psmallmatrix}$ and $\begin{psmallmatrix}
1&1\\ 1&1
\end{psmallmatrix}$. The piece $F^{2,2}/F^{3,1}$ is given by the sum of the four entries. 
In the interpretations below, we consider the columns of the matrices to correspond to O2-planes, when we interpret our
M-theory configuration in type IIA by shrinking one cycle of the torus (as we can do for $k=2$). 
Of course everything should work out correctly in this $k=2$ case, and has been already worked out in \cite{Hanany:2000fq,Hanany:2001iy}. 
We reproduce the analysis here since it is a useful warm-up for $k=3,4,6$.

\paragraph{Shift $=0/2 \in F^{2,2}/F^{3,1}$:}
\[
\begin{array}{c|ccccccc}
\hline
\text{label}& && \begin{psmallmatrix}
0&0\\ 0 &0
\end{psmallmatrix} & \begin{psmallmatrix}
1&1\\ 1&1
\end{psmallmatrix} \\
\text{IIA} &&& O2{}^- + O2{}^- & \widetilde{O2}{}^-+\widetilde{O2}{}^- \\
\text{\#D3} &&&-\frac14 &-\frac14+1 \\
\hline
\text{label}&  \begin{psmallmatrix}
0&1\\ 0 &1
\end{psmallmatrix}& \begin{psmallmatrix}
1&0\\ 1&0
\end{psmallmatrix} && &
\begin{psmallmatrix}
0&0\\ 1 &1
\end{psmallmatrix} &
\begin{psmallmatrix}
1&1\\ 0&0
\end{psmallmatrix}
 \\
\text{IIA}  & O2{}^- +\widetilde{O2}{}^-   &\widetilde{O2}{}^- + O2{}^-  &&& O2{}^+ +O2{}^+ & \widetilde{O2}{}^++\widetilde{O2}{}^+\\
\text{\#D3} &+\frac14 &+\frac14 && & +\frac14 &+\frac14\\
\hline
\text{label}& && \begin{psmallmatrix}
0&1\\ 1 &0
\end{psmallmatrix} & \begin{psmallmatrix}
1&0\\ 0&1
\end{psmallmatrix} \\
\text{IIA}&&& O2{}^+ + \widetilde{O2}{}^+ & \widetilde{O2}{}^++{O2}{}^+ \\
\text{\#D3} &&&+\frac14 &+\frac14 \\
\hline
\end{array}
\]

The first two come  from $O3{}^-$, but the latter of the two has one additional mobile D3-brane stuck at the origin, due to a non-trivial Wilson line  around $S^1$ in the component of $\mathrm{O}(2N)$ disconnected from the identity.
The rest all come from the other three O3-planes wrapped on $S^1$. 

\paragraph{Shift $=1/2 \in F^{2,2}/F^{3,1}$:}

\[
\begin{array}{c|ccccccc}
\hline
\text{label}& && \begin{psmallmatrix}
1&0\\ 0 &0
\end{psmallmatrix} & \begin{psmallmatrix}
0&1\\ 1&1
\end{psmallmatrix} \\
\text{IIA} &&& \widetilde{O2}{}^+ + O2{}^- & {O2}{}^-+\widetilde{O2}{}^- \\
\text{\#D3} &&&0 &0+\frac12 \\
\hline
\text{label}&  \begin{psmallmatrix}
1&1\\ 0 &1
\end{psmallmatrix}& \begin{psmallmatrix}
0&0\\ 1&0
\end{psmallmatrix} && &
\begin{psmallmatrix}
1&0\\ 1 &1
\end{psmallmatrix} &
\begin{psmallmatrix}
0&1\\ 0&0
\end{psmallmatrix}
 \\
\text{IIA}  & \widetilde{O2}{}^+ +\widetilde{O2}{}^-   &{O2}{}^+ + O2{}^-  &&& \widetilde{O2}{}^- +O2{}^+ & {O2}{}^-+\widetilde{O2}{}^+\\
\text{\#D3} &0+\frac12 &0 && & 0 &0+\frac12\\
\hline
\text{label}& && \begin{psmallmatrix}
1&1\\ 1 &0
\end{psmallmatrix} & \begin{psmallmatrix}
0&0\\ 0&1
\end{psmallmatrix} \\
\text{IIA}&&& \widetilde{O2}{}^- + \widetilde{O2}{}^+ & {O2}{}^-+{O2}{}^+ \\
\text{\#D3} &&&0+\frac12 &0\\
\hline
\end{array}
\]
They are
 $\bZ_2$ shift-orientifolds around $S^1$.
More precisely, the geometry is a $\bC^3$ fibration over $S^1$ such that when we go around $S^1$, we have a multiplication by $(-1)$ on $\bC^3$.
The ones with D3-brane charge $0$ are an empty shift-orientifold, and the ones with charge $0+1/2$ have one D3-brane wrapped around $S^1$. 
Note that the charge $1/2$ we are seeing here reflects the fact that the $T^2$ of M-theory is fibered over $\bC^3/\bZ_2$. 
In the type IIB frame, a fiber over a particular point on $S^1$ is $\bC^3$, and therefore the D3-brane charge is 1 if we integrate over the asymptotic infinity of this $\bC^3$.

\subsubsection{$k=3$}\label{sec:k=3}
Let us denote the discrete charges of the three fixed points as elements in $(\bZ_3,\bZ_3,\bZ_3)$. The piece $F_{4,0}=\bZ_3$ consists of $(0,0,0)$, $(1,1,1)$, $(2,2,2)=-(1,1,1)$.
 The piece $F^{2,2}/F^{3,1}=\bZ_3$ corresponding to the shift is given by the sum of the entries. 
 There are 27 choices in total.

\paragraph{Shift $=0/3 \in F^{2,2}/F^{3,1}$:} 
\[
\begin{array}{c|ccc}
\hline
\text{label} &(0,0,0)&  (1,1,1)&  (2,2,2) \\
\text{\#D3} & -\frac13&  -\frac13+1&  -\frac13+1 \\
\hline
\text{label} &  (0,1,2)&   (1,2,0)&   (2,0,1)  \\
\text{\#D3}& +\frac13 &+\frac13&+\frac13\\
\hline
\text{label}& (0,2,1)&   (1,0,2)&   (2,1,0)  \\
\text{\#D3}& +\frac13 &+\frac13&+\frac13\\
\hline
\end{array}
\]

Among the first three, the first entry has the right D3-brane charge to be the  $(k=3,\ell=1)$ S-fold, see  \eqref{variants}. The other two have one more mobile D3-brane, stuck at the origin through a non-trivial $\bZ_3$ background holonomy around $S^1$.

The others all have the right D3-charge to be the $(k=3,\ell=3)$ S-fold, see  \eqref{variants} again.
The second three and the third three have opposite $\bZ_3$ charge characterizing the type of the S-fold.

\paragraph{Shift $=1/3 \in F^{2,2}/F^{3,1}$:} 
\[
\begin{array}{c|ccc}
\hline
\text{label} &(1,0,0)&  (2,1,1)&  (0,2,2) \\
\text{\#D3} & 0&  \frac23&  \frac13\\
\hline
\text{label} &  (1,1,2)&   (2,2,0)&   (0,0,1)  \\
\text{\#D3}& \frac23 &\frac13&0\\
\hline
\text{label}& (1,2,1)&   (2,0,2)&   (0,1,0)  \\
\text{\#D3}& \frac23 &\frac13&0\\
\hline
\end{array}
\]

They are  the $\bZ_3$ shift-S-folds, with the rotation angle $2\pi/3$, of the flat background on $S^1$, with 0, 1, 2  D3-brane(s) wrapped around at the origin.

\paragraph{Shift $=2/3 \in F^{2,2}/F^{3,1}$:} 
\[
\begin{array}{c|ccc}
\hline
\text{label} &(2,0,0)&  (0,1,1)&  (1,2,2) \\
\text{\#D3} & 0&  \frac23&  \frac13\\
\hline
\text{label} &  (2,1,2)&   (0,2,0)&   (1,0,1)  \\
\text{\#D3}& \frac23 &0 & \frac13\\
\hline
\text{label}& (2,2,1)&   (0,0,2)&   (1,1,0)  \\
\text{\#D3}& \frac23 &0&\frac13\\
\hline
\end{array}
\]

They are again the $\bZ_3$ shift-S-folds, but with the rotation angle $4\pi/3$,  of the flat background on $S^1$, with 0, 1, 2 mobile  D3-brane(s) wrapped around at the origin.

\subsubsection{$k=4$}
We denote the charges at the fixed points as elements in $(\bZ_4,\bZ_4;\bZ_2)$.
The piece $F_{4,0}=\bZ_4$ consists of $(0,0;0)$, $(1,1;1)$, $(2,2;0)$, $(3,3;1)$.

\paragraph{Shift $=0/4 \in F^{2,2}/F^{3,1}$:} 
\[
\begin{array}{c|cccc}
\hline
\text{label} &  (0,0;0)&   (1,1;1)&   (2,2;0)&   (3,3;1)  \\
\text{\#D3} &  -\frac38&   -\frac38+1&   -\frac38+1&   -\frac38+1 \\
\hline
\text{label} &  (0,2;1)&   (1,3;0)&   (2,0;1)&   (3,1;0)  \\
\text{\#D3} & +\frac38 & +\frac38 & +\frac38 & +\frac38\\
\hline
\end{array}
\]
Among  the first four, the first has the right D3-brane charge to be the  $(k=4,\ell=1)$ S-fold, see  \eqref{variants}.
The other three have one more mobile D3-brane, stuck at the origin through non-trivial $\bZ_4$ background holonomy around $S^1$.
All the second four have the right D3-charge to be the $(k=4,\ell=4)$ S-fold, see \eqref{variants}.

\paragraph{Shift $=1/4 \in F^{2,2}/F^{3,1}$:} 
\[
\begin{array}{c|cccc}
\hline
\text{label} &  (0,1;0)&   (1,2;1)&   (2,3;0)&   (3,0;1)  \\
\text{\#D3} &  0&   +\frac34&   +\frac12&   +\frac14  \\
\hline
\text{label} &  (0,3;1)&   (1,0;0)&   (2,1;1)&   (3,2;0)  \\
\text{\#D3} & +\frac14&   0&   +\frac34&   +\frac12 \\
\hline
\end{array}
\]
These are $\bZ_4$  shift-S-folds of flat space, with rotation angle $\pi/2$,
and with $0$, $1$, $2$ or $3$ mobile D3-brane(s) around $S^1$.

\paragraph{Shift $=2/4 \in F^{2,2}/F^{3,1}$:} 
\[
\begin{array}{c|cccc}
\hline
\text{label} &  (0,0;1)&   (1,1;0)&   (2,2;1)&   (3,3;0)  \\
\text{\#D3} & -\frac18&   -\frac18+\frac12&   -\frac18+1&   -\frac18+\frac12  \\
\hline
\text{label} &  (0,2;0)&   (1,3;1)&   (2,0;0)&   (3,1;1)  \\
\text{\#D3} &  +\frac18&   +\frac18+\frac12&   +\frac18&   +\frac18+\frac12 . \\
\hline
\end{array}
\]
These are the $\bZ_2$ shift-S-folds  on $S^1$ of the O3-planes. 

Among the first four, the first has the right D3-brane charge to be the $\bZ_2$ shift-S-fold of $O3{}^-$.
Recall that the base of the M-theory configuration is $\bC^3/\bZ_4$, but the in the type IIB description, the fiber at a particular point on $S^1$ is $\bC^3/\bZ_2$. The D3-brane charge of $O3{}^-$ is $-1/4$, and therefore the charge as seen from the M-theory configuration is $1/2\cdot (-1/4)=-1/8$.

The others have one or two additional mobile D3-brane(s) trapped at the origin.
Among the second four, the first and the third have the right D3-brane charge to be the $\bZ_2$ shift-S-fold of $\widetilde{O3}{}^+$.
The second and the fourth have one additional mobile D3-brane trapped at the origin.

Note that we cannot take the $\bZ_2$ shift S-fold of $\widetilde{O3}{}^-$ or $O3{}^+$,
since they are exchanged by this $\bZ_2$ operation. 

\paragraph{Shift $=3/4 \in F^{2,2}/F^{3,1}$:} 
\[
\begin{array}{c|cccc}
\hline
\text{label} &  (0,1;1)&   (1,2;0)&   (2,3;1)&   (3,0;0)  \\
\text{\#D3} &+\frac14&   +\frac12&   +\frac34&   0   \\
\hline
\text{label} &  (0,3;0)&   (1,0;1)&   (2,1;0)&   (3,2;1)  \\
\text{\#D3} &  0&   +\frac14&   +\frac12&   +\frac34 .\\
\hline
\end{array}
\]
These are $\bZ_4$  shift-S-folds of flat space, with rotation angle $3\pi/2$,
and with $0$, $1$, $2$ or $3$ mobile D3-brane(s) around $S^1$.

\subsubsection{$k=6$}
We denote the charges at the fixed points as elements in $(\bZ_6,\bZ_3,\bZ_2)$.
The piece $F_{4,0}=\bZ_6$ consists of $(0,0,0)$, $(1,1,1)$, $(2,2,0)$, $(3,0,1)$, $(4,1,0)$, $(5,2,1)$.

\paragraph{Shift $=0/6 \in F^{2,2}/F^{3,1}$:} 
\[
\begin{array}{c|cccccc}
\hline
\text{label}&  (0,0,0)&   (1,1,1)&   (2,2,0)&   (3,0,1)&   (4,1,0)&   (5,2,1)  \\
\text{\#D3}& -\frac5{12}&-\frac5{12}+1&-\frac5{12}+1&-\frac5{12}+1&-\frac5{12}+1&-\frac5{12}+1\\
\hline
\end{array}
\]
The first has the right D3-brane charge to be the  $(k=6,\ell=1)$ S-fold, see  \eqref{variants}.
The other five have one more mobile D3-brane, stuck at the origin through a non-trivial $\bZ_6$ background holonomy around $S^1$.

\paragraph{Shift $=1/6 \in F^{2,2}/F^{3,1}$:} 
\[
\begin{array}{c|cccccc}
\hline
\text{label}&   (1,0,0)&   (2,1,1)&   (3,2,0)&   (4,0,1)&   (5,1,0)&   (0,2,1)   \\
\text{\#D3}&  0&   \frac56&   \frac23&   \frac12&   \frac13&   \frac16 \\
\hline
\end{array}
\]
This is the $\bZ_6$ shift-S-fold of flat space, with zero to five D3-branes stuck at the origin. The rotation angle is $2\pi/6$.

\paragraph{Shift $=2/6 \in F^{2,2}/F^{3,1}$:} 
\[
\begin{array}{c|cccccc}
\hline
\text{label}&   (2,0,0)&   (3,1,1)&   (4,2,0)&   (5,0,1)&   (0,1,0)&   (1,2,1)  \\
\text{\#D3}&  -\frac1{12}+\frac13&   -\frac1{12}+1&   -\frac1{12}+\frac23&   -\frac1{12}+\frac13&   -\frac1{12}&   -\frac1{12}+\frac23 \\
\hline
\end{array}
\]
The fifth has the correct charge to be the $\bZ_3$ shift-S-fold of the standard $O3{}^-$-plane. 
Since the $O3{}^-$-plane itself has an identification by the angle $\pi$, its $\bZ_3$ quotient involves the rotation by $\pi/3$ of the transverse space.
Note also that the $O3{}^-$-plane has the D3-brane charge $-1/4$, therefore we see $1/3\cdot(-1/4)=-1/12$ in M-theory.
The others have one or two additional D3-brane(s) at the origin.

\paragraph{Shift $=3/6 \in F^{2,2}/F^{3,1}$:} 
\[
\begin{array}{c|cccccc}
\hline
\text{label}&   (3,0,0)&   (4,1,1)&   (5,2,0)&   (0,0,1)&   (1,1,0)&   (2,2,1)  \\
\text{\#D3}&  -\frac16+\frac12&   -\frac16+1&   -\frac16+\frac12&   -\frac16&   -\frac16+\frac12&   -\frac16+1 \\
\hline
\end{array}
\]
The fourth has the correct charge to be the $\bZ_2$ shift-S-fold of the $(k=3,\ell=1)$ S-fold. The others have one or two additional mobile D3-brane(s) on top.

\paragraph{Shift $=4/6 \in F^{2,2}/F^{3,1}$:} 
\[
\begin{array}{c|cccccc}
\hline
\text{label}&  (4,0,0)&   (5,1,1)&   (0,2,0)&   (1,0,1)&   (2,1,0)&   (3,2,1)  \\
\text{\#D3}&  -\frac1{12}+\frac13&   -\frac1{12}+\frac23&   -\frac1{12}&   -\frac1{12}+\frac13&   -\frac1{12}+\frac23&   -\frac1{12}+1 \\
\hline
\end{array}
\]

The third has the correct charge to be the $\bZ_3$ shift-S-fold, of rotation angle $2\pi/3$, of the standard $O3{}^-$-plane. The others have one or two additional D3-brane(s) at the origin.

\paragraph{Shift $=5/6 \in F^{2,2}/F^{3,1}$:} 
\[
\begin{array}{c|cccccc}
\hline
\text{label}&  (5,0,0)&   (0,1,1)&   (1,2,0)&   (2,0,1)&   (3,1,0)&   (4,2,1)  \\
\text{\#D3}& 0&   \frac16&   \frac13&   \frac12&   \frac23&   \frac56  \\
\hline
\end{array}
\]

This is the $\bZ_6$ shift-S-fold of flat space, with zero to five D3-branes stuck at the origin. The rotation angle is $2\pi \cdot 5/6$.

\paragraph{Comments:}
Note that in the cases with shift $2,4\in \bZ_6$, we only find  $\bZ_3$ shift-S-folds of ${O3}{}^-$, but we
 \emph{do not} have $\bZ_3$ shift-S-folds of $\widetilde{O3}{}^-$, $O3{}^+$ and $\widetilde{O3}{}^+$.  This is as it should be, because these three types of O3-planes  are permuted by the $\bZ_3$ action.

Similarly, there is no $\bZ_2$ shift-orientifold of the $(k=3,\ell=3)$ orientifold, since the $\bZ_2$ action exchanges the two subtly different versions that we denoted by $(1,1,1)$ and (2,2,2) in Sec.~\ref{sec:k=3}.

\section{Special cases and \Nequals4 enhancement}
\label{sec5}
So far, we saw that $N$ D3-branes probing various variants of the S-folds give rise to \Nequals3 superconformal field theories characterized by 
\begin{equation}
(k,\ell)=(3,1), (3,3), (4,1), (4,4), (6,1).
\end{equation}
When $\ell < k$ we expect to have both `parent' theories and their discrete gaugings. In this section we discuss some interesting special cases, including cases where the ${\cal N}=3$ supersymmetry is enhanced to ${\cal N}=4$.


As discussed in \cite{Aharony:2015oyb}, an enhancement of supersymmetry to \Nequals4 occurs if and only if there is a Coulomb branch operator of dimension 1 or 2, since \Nequals3 supersymmetry then dictates the presence of extra supercharges.
The dimensions of the Coulomb branch operators of our `parent' \Nequals3 theories were given in \eqref{dims}.
For $k=2$ we always have such an enhancement, but for $k=3,4,6$ we see that it happens just for $\ell=1$ and $N=1,2$.
Note that the theory with the lowest central charge which does not have any enhancement is the $N=1$, $k=\ell=3$ theory, whose only Coulomb branch operator has dimension 3. The central charges of this theory are the same as those of five vector multiplets. 
Since the Coulomb branch operators of \Nequals3 theories must be integers as shown in Section 3.1 of \cite{Nishinaka:2016hbw}, then this must be the `minimal' \Nequals3 SCFT, assuming the general validity of the formula \eqref{2a-c}.  It would be interesting to test this by a superconformal bootstrap analysis, generalizing the ones in \cite{Beem:2013qxa,Liendo:2015ofa}.

Going back to theories with \Nequals4 supersymmetry, the case $N=1$  is rather trivial: we just have a Coulomb branch operator of dimension one, so the moduli space is just $\bC^3$, and we get the \Nequals4 super Yang-Mills theory with gauge group $\U(1)$. 
So let us discuss the $N=2$ cases. 

The spectrum of the Coulomb branch operators of the `parent' theory is given by \begin{equation}
\begin{cases}
2,3 & (k=3) ,\\
2,4 & (k=4) ,\\
2,6 & (k=6) .
\end{cases}
\end{equation}
These spectra agree with those of an \Nequals4 super Yang-Mills theory with gauge group $\SU(3)$, $\SO(5)$ and $G_2$, respectively.
Below we give evidence that indeed, the `parents' of these S-fold configurations give rise to these \Nequals4 super Yang-Mills theories, realized in a somewhat unusual manner. Note that ${\cal N}=4$ theories always have an exactly marginal deformation, sitting in the same multiplet as the dimension two Coulomb branch operator, and our conjectured relation implies that for our $N=2$ theories this is the gauge coupling of these ${\cal N}=4$ gauge theories. Our discussion in the previous sections implies that in the AdS dual of these `parent' theories, this marginal deformation corresponds to a scalar field coming from a D3-brane wrapped on the torsion 3-cycle; for this specific case this wrapped D3-brane gives rise to a massless field.

Again let us limit ourselves to the points on the moduli space corresponding to a Coulomb branch from the point of view of an ${\cal N}=2$ description of our SCFTs. In the $N=2, \ell=1$ theories, this subspace is parameterized by $z_{1,2}\in \bC$, with
the gauge symmetry \eqref{newident} 
\begin{equation}
(z_1,z_2) \mapsto (\gamma^n z_2,\gamma^{-n} z_1), \label{gauge}
\end{equation} where $\gamma=e^{2\pi i/k}$ and $n$ is any integer. 
The charges in a basis that is natural from this point of view can be written as $(e_1,m_1;e_2,m_2)$. 

Naively, one would expect the electric charges $e_1,e_2$ to correspond to electric charges of the corresponding ${\cal N}=4$ theory, but this cannot be the case because of the non-trivial $Sp(4,\bZ)$ action on these charges, induced by the $\SL(2,\bZ)$ transformation that accompanies the identification \eqref{gauge}. So instead we consider the
rank-2 sublattice containing charges of the form $(Q;\bar Q)$ where we regard $(e,m)\in \bZ\oplus\bZ $ as a complex number $Q = e + m \gamma$.

Two charges from this sublattice are local with respect to each other. To see this, note that given $Q=e+m\gamma$ and $Q'=e'+m'\gamma$, their Dirac pairing is \begin{equation}
em'-e'm = (\bar Q Q'-\bar Q{}' Q)/(\gamma-\bar\gamma).
\end{equation}  Then the Dirac pairing between the two charges $(Q;\bar Q)$ and $(Q';\bar Q{}')$ is clearly zero.
This means that we can take these charges to be the ``electric charges'' in the \Nequals4 description. We provide some consistency checks for this below. 

Now, note that the $\SL(2,\bZ)$ element associated to the $\bZ_k$ orbifold then acts on $Q$ just by multiplication by $\gamma$.
Then the gauge transformation \eqref{gauge} acts on this variable $Q$ as \begin{equation}
Q \mapsto \gamma^n\bar Q,
\end{equation} which is a reflection of the complex plane along the line  $e^{\pi in/ k}\bR$.
This makes it clear that the group generated by \eqref{gauge} for $k=3,4,6$ is the Weyl group of $\SU(3)$, $\SO(5)$ and $G_2$, respectively.

Let us test our identification by looking at half-BPS particles. There's no string connecting $z_1$ and $\gamma^n z_1$, since we know nothing happens when $z_1=0$ for $\ell=1$. So there are only strings connecting $\gamma^n z_1$ and $\gamma^m z_2$. 
Using \eqref{gauge} we can always restrict $z_1$ to have a phase between 0 and $2\pi/k$. 
Then we just have to consider all $(p,q)$-strings connecting $z_1$ and $\gamma^n z_2$ for $n=0,\ldots,k-1$.  
Since the IIB coupling constant is $\tau=\gamma$, the central charges of half-BPS particles are given by 
\begin{equation}
(p+q\gamma)(z_1 - \gamma^n z_2).\label{cc}
\end{equation} 

We conjectured that ``electric'' states have the charge $(Q;\bar Q)$, and then their central charges are given by $Qz_1-\bar Q z_2$. Comparing with \eqref{cc}, we see that ``electric'' objects have the following $Q$:
\begin{itemize}
\item For $k=3$, $Q=\omega^n$ and $Q=(1+\omega)\omega^n=-\omega^{n-1}$ where $\omega=e^{2\pi i/3}$,
\item For $k=4$, $Q=i^n$ and $Q=(1+i) i^n$,
\item For $k=6$, $Q=\gamma^n$ and $Q=(1+\gamma)\gamma^n$.
\end{itemize}
Clearly they can be identified with the roots of $\SU(3)$, $\SO(5)$, and $G_2$, respectively. 

The metric on the moduli space is also correctly mapped to that on the Cartan subalgebra of these groups.
The original metric is $dz_1 d\bar z_1 + dz_2 d\bar z_2$ on $\bC^2$,
and  we choose a real subspace $\bR^2$ of the form $Qz_1-\bar Q z_2$.
Then, two vectors $Qz_1-\bar Qz_2$ and  $Q' z_1-\bar Q'z_2$ in $\bR^2$ have the induced inner product $(2\mathrm{Re}(Q\bar Q'))$.
Using this, we can easily check that the vectors listed above have the same inner products as the root vectors of $\SU(3)$, $\SO(5)$, and $G_2$, in the normalization that the short roots have length squared 2.

Finally, recall that the dyons of \Nequals4 SYM have central charges of the form \begin{equation}
(p+q \tau_{YM}) (\alpha_s\cdot \phi), \qquad (p+q \frac{\tau_{YM}}{r}) (\alpha_l \cdot \phi),\label{n4}
\end{equation} where $\alpha_{s,l}$ are short and long roots, and $r$ is the length squared of the long roots divided by that of the short roots. 
We can check that the spectrum \eqref{cc} can be matched with  \eqref{n4} with the identification of the roots given above, if we take $\tau_{YM}=-1/(1+\gamma)$, uniformly for $k=3,4,6$.
As a further check, note that for $k=6$ this is exactly the value of $\tau_{YM}$ for which the $G_2$ ${\cal N}=4$ theory has a discrete $\bZ_6$ symmetry \cite{Argyres:2006qr}.


\section*{Acknowledgements}
The authors would like to thank I. Garc\'\i a-Etxebarria, D. Harlow, Z. Komargodski, T. Nishinaka, H. Ooguri and N. Seiberg for useful discussions.
The work of OA is supported in part by an Israel Science Foundation center for excellence grant, by the I-CORE program of the Planning and Budgeting Committee and the Israel Science Foundation (grant number 1937/12), by the Minerva foundation with funding from the Federal German Ministry for Education and Research, by a Henri Gutwirth award from the Henri Gutwirth Fund for the Promotion of Research, and by the ISF within the ISF-UGC joint research program framework (grant no. 1200/14). OA is the Samuel Sebba Professorial Chair of Pure and Applied Physics.
The work of YT is partially supported in part by JSPS Grant-in-Aid for Scientific Research No. 25870159,
and  by WPI Initiative, MEXT, Japan at IPMU, the University of Tokyo.

\appendix

\section{Properties of cohomology groups relevant for S-folds}
\label{appendix}


In this appendix\footnote{%
Contributed by Kiyonori Gomi in 2023,
who noticed a question posted by Y. Tachikawa on MathOverflow at \url{https://mathoverflow.net/questions/229700/}.
}, we prove the properties \eqref{aaaa} and \eqref{bbbb} 
concerning the cohomology groups associated to the S-folds
relevant in their understanding.
The appendix is written in the rigorous mathematical style,
and can be read independently.
It includes a brief review of Borel equivariant cohomology,
which is heavily used in the following proofs.

\subsection{The setup and the propositions to be proved}

Let $\Z_k = \{ u \in U(1) |\ u^k = 1 \}$ be the cyclic group of order $k$, generated by $\zeta = \exp 2\pi \sqrt{-1}/k$. In the note, the order will be $k = 2, 3, 4, 6$. We let $\Z_k$ act on the complex vector space $\C^3$ and the two-dimensional torus $T^2 = \C/(\Z \oplus \Z \tau_k)$ by
\begin{align*}
\C^3 &\to \C^3, & 
(z_1, z_2, z_3) &\mapsto (\zeta z_1, \zeta z_2, \zeta z_3), \\
T^2 &\to T^2, &
z &\mapsto \zeta z,
\end{align*}
where $\tau_2 = \tau_4 = \sqrt{-1}$, $\tau_3 = \tau_6 = 2\pi \sqrt{-1}/6$. We define
$$
X := (D^6 \times T^2)/\Z_k,
$$
where $\Z_k$ acts on both the unit disk $D^6 \subset \C^3$ and $T^2$.

The action of $\Z_k$ on $\widetilde{X} := D^6 \times T^2$ is not free. We write for $\widetilde{\Sing}$ for the set of points which have non-trivial stabilizer groups 
$$
\widetilde{\Sing} := \{ x \in \widetilde{X} |\ 
\exists u \in \Z_k \backslash \{ 1 \}, ux = x \}.
$$
The image $\Sing = \pi(\widetilde{\Sing}) \subset X$ under the natural projection $\pi : \widetilde{X} \to X$ is the set of orbifold singularities. We choose a small closed disk $D^8(p) \subset X$ around each point $p \in \widetilde{\Sing}$, so that their (disjoint) union $D^8(\widetilde{\Sing}) = \sqcup_p D^8(p)$ becomes $\Z_k$-invariant. Removing the interior of $D^8(\widetilde{\Sing})$, we define
\begin{align*}
\widetilde{X}' &:= \widetilde{X} - \mathrm{Int}(D^8(\widetilde{\Sing})), &
X' &:= \widetilde{X}'/\Z_k = X - \mathrm{Int}(D^8(\widetilde{\Sing})/\Z_k).
\end{align*}

The action of $\Z_k$ on $\widetilde{X}'$ is free. Hence $X'$ is a $8$-dimensional manifold, whose boundary $\partial X' = Y_{\mathrm{in}} \sqcup Y_{\mathrm{out}}$ has two types of boundary components
\begin{align*}
Y_{\mathrm{in}} &:= \partial D^8(\widetilde{\Sing})/\Z_k, &
Y_{\mathrm{out}} &:= \partial \widetilde{X}/\Z_k
= (S^5 \times T^2)/_{\Z_k}.
\end{align*}
The ``inner boundary'' $Y_{\mathrm{in}}$ turns out to be a disjoint union of $7$-dimensional lens spaces, and the ``outer boundary'' $Y_{\mathrm{out}}$ is a fiber bundle over the $5$-dimensional lens space $S^5/\Z_k$ with fiber $T^2$.


In this appendix, 
we first study the precise relationship between the $4$th cohomology of $Y_{\mathrm{in}}$ and $Y_{\mathrm{out}}$ by restricting that of $X'$:

\begin{prop} \label{prop:restriction_is_bijective}
The inclusions $Y_{\mathrm{in}}, Y_{\mathrm{out}} \subset X'$ induce isomorphisms of cohomology groups
$$
H^4(Y_{\mathrm{in}}; \Z) \cong
H^4(X'; \Z) \cong
H^4(Y_{\mathrm{out}}; \Z).
$$
\end{prop}

To Proposition \ref{prop:restriction_is_bijective}, the key idea may be the use of the Borel equivariant cohomology: Since $Y_{\mathrm{out}} = (S^5 \times T^2)/\Z_k$ is the quotient space under a free action, we have
$$
H^n(Y_{\mathrm{out}}; \Z)
= H^n((S^5 \times T^2)/\Z_k); \Z)
\cong H^n_{\Z_k}(S^5 \times T^2; \Z).
$$
Then, regarding $S^5 \times T^2$ as a $S^5$-bundle over $T^2$, rather than a $T^2$-bundle over $S^5$, we can apply the Gysin exact sequence (the Leray-Serre spectral sequence for sphere bundles), which is easier to compute in the present case. The relationship of the cohomology groups of $X'$, $Y_{\mathrm{in}}$ and $Y_{\mathrm{out}}$ is then studied by using the Mayer-Vietoris sequence computing the equivariant cohomology of $D^6 \times T^2 \simeq T^2$.

\bigskip

The next question is the relationship between the graded quotient of the Leray-Serre spectral sequence for $T^2 \to Y_{\mathrm{out}} \to S^5/\Z_k$ and $H^4(Y_{\mathrm{in}}; \Z)$. Let $F^p = F^{p, 4-q} = F^{p, 4-p}H^4(Y_{\mathrm{out}}; \Z)$ be the filtration in the spectral sequence, which reads
$$
H^4(Y_{\mathrm{out}}; \Z) = F^0 = F^1 = F^2
\supset F^3 \supset F^4 \supset F^5 = 0
$$
in the present case. Then there naturally arise the following homomorphisms
$$
F^4 \to H^4(Y_{\mathrm{out}}; \Z) \to F^2/F^3.
$$
By Proposition \ref{prop:restriction_is_bijective}, we can identify $H^4(Y_{\mathrm{out}}; \Z)$ with $H^4(Y_{\mathrm{in}}; \Z)$ in a natural way, so that they can be thought of as ``the same''. Under this identification, the following proposition establishes the properties \eqref{aaaa} and \eqref{bbbb} in the main text, which were conjectured from 
a specific case of $k=2$ and the consistency with expected behaviors of the duality between M-theory and F-theory:

\begin{prop} \label{prop:maps}
For any $k = 2, 3, 4, 6$, the following holds true.
\begin{itemize}
\item[(a)]
$F^4 \to H^4(Y_{\mathrm{out}}; \Z)$ is the `diagonal embedding'.

\item[(b)]
$H^4(Y_{\mathrm{out}}; \Z) \to F^2/F^3$ is the `sum'.

\end{itemize}
\end{prop}

The meaning of the `diagonal embedding' and the `sum' will be given in the body of the appendix,
which needs a more explicit description of the relevant groups.

Proposition \ref{prop:maps} is also established by using the $\Z_k$-equivariant cohomology of $T^2$. If one notices that the classifying space $B\Z_k$ of $\Z_k$ is the colimit of the odd dimensional lens spaces $S^{2r-1}/\Z_k$, the filtration of the cohomology of $Y_{\mathrm{out}} = (S^5 \times T^2)/\Z_k$ is compatible with a filtration of the equivariant cohomology of $T^2$. Then Proposition \ref{prop:maps} will follow from the corresponding claims in the equivariant cohomology. The counterpart of (a) is rather easy to show. The counterpart of (b) is shown by identifying the homomorphism in question with the push-forward map along the map $T^2 \to \pt$.

\bigskip

The rest of the appendix is organized as follows. In Subsection \ref{sec:equivariant_coh}, a brief review about the Borel equivariant cohomology is supplied. Then in Subsection \ref{sec:proof}, the propositions will be proved.

\medskip

As a notation, we use both $\Z_k$ and $\Z/k$ to mean the cyclic group of order $k$. The distinction is that the group multiplication in the former $\Z_k = \{ u \in U(1) |\ u^k = 1 \}$ is multiplicative, and that in the latter $\Z/k$ is additive.


\subsection{Borel equivariant cohomology}
\label{sec:equivariant_coh}

This section is a short review of the Borel equivariant cohomology.

\subsubsection{Definition}

Let $G$ be a compact Lie group acting on a space $X$. We assume implicitly that $X$ is nice enough. Then the $G$-equivariant cohomology of $X$ with coefficients in $\Z$ (in the sense of Borel) is defined as
$$
H^n_G(X; \Z) := H^n(EG \times_G X; \Z),
$$
where $n \in \Z$, and $EG$ is the total space of the universal bundle over the classifying space $BG$ of $G$. The action of $g \in G$ on $(\xi, x) \in EG \times X$ is given by $g(\xi, x) = (\xi g^{-1}, gx)$. The quotient space $EG \times_G X$ is often called the homotopy quotient. In the same way, the relative cohomology $H^n_G(X, Y; \Z)$ is defined for any subspace $Y \subset X$ which is invariant in the sense that $gy \in Y$ for all $g \in G$ and $y \in Y$. Accordingly, if $\pt \in X$ is a fixed point, $g \cdot \pt = \pt$ for all $g \in G$, then the reduced cohomology is defined as 
$$
\tilde{H}^n_G(X; \Z) := H^n_G(X, \pt; \Z).
$$
As in the non-equivariant case, we have the direct sum decomposition
$$
H^n_G(X; \Z) \cong H^n_G(\pt; \Z) \oplus \tilde{H}^n_G(X; \Z).
$$

\bigskip

By the cup product, the equivariant cohomology forms the equivariant cohomology ring $$H^*_G(X; \Z) = \bigoplus_{n \in \Z} H^n_G(X; \Z).$$ There always exists the trivial $G$-equivariant map $X \to \pt$, so that we have an induced ring homomorphism $H^*_G(\pt; \Z) \to H^*_G(X; \Z)$. Through this homomorphism, the equivariant cohomology of $X$ is regarded as a module over the ring $H^*_G(\pt; \Z) \cong H^*(BG; \Z)$.

\bigskip

As is well-known, if $G$ acts on $X$ freely, then the Borel equivariant cohomology is naturally isomorphic to the non-equivariant cohomology of the quotient space
$$
H^n_G(X; \Z) \cong H^n(X/G; \Z).
$$
This is induced from the the projection $EG \times_G X \to X/G$, ($[\xi, x] \mapsto [x]$). If $H \subset G$ is a subgroup and $G$ acts on $G/H$ by the multiplication in $G$, then there is also an isomorphism $H^n_G(G/H; \Z) \cong H^n_H(\pt; \Z)$ for all $n \in \Z$. This is induced from the inclusion $\pt \to G$ to the unit element in $G$.

\subsubsection{Gysin exact sequence}

Let $G$ be a compact Lie group, and $\pi : E \to X$ a $G$-equivariant complex vector bundle of (complex) rank $r$.  We assume that $E$ is equipped with a $G$-invariant Hermitian metric. Write $D(E) \subset E$ for the unit disk bundle of $E$, and $S(E) \subset D(E)$ for the unit sphere bundle. Rewriting the Thom isomorphism for the (non-equivariant) complex vector bundle $EG \times_G E \to EG \times_G X$, we get the Thom isomorphism
$$
H^n_G(X; \Z) \cong H^{n + 2r}_G(D(E), S(E); \Z).
$$
Note that the projection $\pi : D(E) \to X$ provides a $G$-equivariant homotopy equivalence $D(E) \simeq X$. Hence $H^n_G(D(E); \Z) \cong H^n(X; \Z)$ for all $n \in \Z$. Substituting this and the Thom isomorphism into the long exact sequence associated to the pair $(D(E), S(E))$, we get the Gysin exact sequence
$$
\cdot\cdot \to
H^{n- 2r}_G(X; \Z) \overset{c^G_r(E)}{\to}
H^n_G(X; \Z) \overset{\pi^*}{\to}
H^n_G(S(E); \Z) \to
H^{n-2r+1}_G(X; \Z) \overset{c^G_r(E)}{\to}
\cdot\cdot.
$$
In the above, $\pi^*$ is the pull-back under the projection $\pi : S(E) \to X$, and $c^G_r(E)$ stands for the cup product with the $r$th $G$-equivariant Chern class of the $G$-equivariant vector bundle $E$
$$
c^G_r(E) := c_r(EG \times_G E) \in
H^{2r}_G(X; \Z).
$$

\bigskip

More generally, the Thom isomorphism exists for $G$-equivariant real vector bundle $\pi : E \to X$ which is $G$-equivariantly oriented. A $G$-equivariant orientation is a trivialization of the $G$-equivariant real line bundle $\det E$. The Thom isomorphism reads
$$
H^n_G(X; \Z) \cong
H^{n + r}_G(D(E), S(E); \Z),
$$
where $r$ is the (real) rank of $E$, and $D(E)$ and $S(E)$ are the disk and sphere bundles associated to a $G$-invariant metric. The Thom isomorphism also leads to the Gysin exact sequence, where the role of the Chern class in the case that $E$ is a $G$-equivariant complex vector bundle is played by the equivariant Euler class $e^G(E) \in H^r_G(X; \Z)$. (Actually, if $E$ is a $G$-equivariant complex vector bundle of complex rank $r$, then $e^G(E) = c_r^G(E)$.)

\bigskip

As an example, let the cyclic group $\Z_k = \{ u \in \C |\ u^k = 1 \}$ of order $k$ act on $\C$ through the inclusion $\Z_k \subset U(1)$ and the scalar multiplication of $U(1) \subset \C$ on $\C$. We can think of $\C$ as a $\Z_k$-equivariant complex vector bundle over a point $\pt$ with the trivial $\Z_k$-action. The Borel equivariant cohomology of $\pt$ is just the ordinary cohomology of $B\Z_k$, which has the ring presentation
\begin{gather*}
H^*_{\Z_k}(\pt; \Z)
= H^*(B\Z_k; \Z)
\cong \Z[t]/( kt ). \\
\begin{array}{|c|c|c|c|c|c|c|c|c|}
\hline
& n = 0 & n = 1 & n = 2 & n = 3 & n = 4 & n = 5 & n = 6 & \cdots \\
\hline
H^n(B\Z_k; \Z) & \Z & 0 & \Z/k & 0 & \Z/k & 0 & \Z/k & \cdots \\
\hline
\mbox{basis} & 1 & & t & & t^2 & & t^3 & \cdots \\
\hline
\end{array}
\end{gather*}
The ring basis $t = c_1^{\Z_k}(\C) \in H^2_{\Z_k}(\pt; \Z) \cong \Z/k$ is nothing but the equivariant first Chern class of the equivariant line bundle associated to the representation $\Z_k \subset U(1)$. For the direct sum $\C^r = \overbrace{\C \oplus \cdots \oplus \C}^r$ of this equivariant line bundle, we have
$$
c_r^{\Z_k}(\C^r) = c_1^{\Z_k}(\C)^r = t^r
\in H^{2r}_{\Z_k}(\pt; \Z) \cong \Z/k.
$$
One can summarize the cohomology groups relevant to the Gysin exact sequence as follows:
$$
\begin{array}{c|c|c|c}
\vdots & \vdots & \vdots & \vdots \\
\hline
n = 2r & \Z/k & \Z/k & \textcolor{red}{0} \\
\hline
n = 2r+ 1 & 0 & 0 & \textcolor{red}{0} \\
\hline
n = 2r & \Z & \Z/k & \textcolor{red}{0} \\
\hline
n = 2r-1 & 0 & 0 & \textcolor{red}{\Z} \\
\hline
n = 2r-2 & 0 & \Z/k & \textcolor{red}{\Z/k} \\
\hline
n = 2n-1 & 0 & 0 & \textcolor{red}{0} \\
\hline
& \vdots & \vdots & \vdots \\
\hline
n = 3 & 0 & 0 & \textcolor{red}{0} \\
\hline
n = 2 & 0 & \Z/k & \textcolor{red}{\Z/k} \\
\hline
n = 1 & 0 & 0 & \textcolor{red}{0} \\
\hline
n = 0 & 0 & \Z & \textcolor{red}{\Z} \\
\hline
& H^{n-2r}_{\Z_k}(\pt; \Z) & H^n_{\Z_k}(\pt; \Z) &
H^n_{\Z_k}(S(\C^d); \Z)
\end{array}
$$
It is then easy to determine the equivariant cohomology of the unit sphere $S(\C^4) \cong S^{2r-1}$, as indicated in the \textcolor{red}{red} colour above. Note that the $\Z_k$-action on this sphere is free, and the quotient is the lens space in dimension $2r-1$. Hence the above calculation determines the ordinary cohomology of the odd dimensional lens space $H^n(S^{2r-1}/\Z_k; \Z) \cong H^n_{\Z_k}(S^{2r-1}; \Z)$. In the relevance to the aim of this appendix, we point out that the pull-back by the projection $\pi : S^{2r-1} \to \pt$ induces an isomorphism
$$
\pi^* : H^n_{\Z_k}(\pt; \Z) \overset{\cong}{\longrightarrow}
H^n_{\Z_k}(S^{2r-1}; \Z) \cong H^n_{\Z_k}(S^{2r-1}/\Z_k; \Z)
$$
in the range $n \le 2r - 2$. Non-equivariantly, this homomorphism is identified with that induced from the inclusion
$$
S^{2r-1}/\Z_k \to \varinjlim_{r} S^{2r-1}/\Z_k = S^\infty/\Z_k = B\Z_k.
$$

\subsubsection{Push-forward}

Let $X$ and $Y$ be closed manifolds with actions of a compact Lie group $G$, and $f : X \to Y$ a $G$-equivariant smooth map. Suppose that $X$ and $Y$ are oriented, and the $G$-actions preserve the orientations. In this setup, we can define the push-forward homomorphism
$$
f_* : H^n_G(X; \Z) \to H^{n - \dim X + \dim Y}_G(Y; \Z)
$$
for all $n \in \Z$. 

\medskip

The homomorphism is defined in a standard manner: By the Mostow-Palais theorem, there is an equivariant embedding $\iota : X \to V$ of $X$ into the representation space $V$ of a real representation of $G$. Without loss of generality, we can assume that the real representation is orientation preserving. We then have an associated equivariant embedding $(f, \iota) : X \to Y \times V$. Let $N \to X$ be the normal bundle of $(f, \iota)$, which is $G$-equivariantly orientable. The total space $N$ is identified with an open tubular neighbourhood of the embedded image of $X$ in $Y \times V$. Let $D(N)$ be the disk bundle of $N$, which is an invariant closed subspace of $Y \times D(V)$, where $D(V)$ is an invariant closed disk of $V$. Now, $f_*$ is defined as the composition of the following homomorphisms.
\begin{align*}
H^n_G(X; \Z) 
&\cong H^{n + \dim Y + \dim V - \dim X}_G(D(N), S(N); \Z) \\
&\quad (\mbox{Thom isomorphism}) \\
&\cong H^{n + \dim Y + \dim V - \dim X}_G
(Y \times D(V), Y \times D(V) - \mathrm{Int}D(N); \Z) \\
&\quad (\mbox{excision axiom}) \\
&\to H^{n + \dim Y + \dim V - \dim X}_G(Y \times D(V), Y \times S(V); \Z) \\
&\quad (\mbox{restriction}) \\
&\cong H^{n + \dim Y - \dim X}_G(Y; \Z). \\
&\quad(\mbox{Thom isomorphism})
\end{align*}

\medskip

The basic properties are as follows:

\begin{enumerate}
\item
(projection formula) For any $x \in H^n_G(X; \Z)$ and $y \in H^m_G(Y; \Z)$, we have $f_*(x \cup f^*(y)) = f_*(x) \cup y$.

\item
If $Z$ is a closed oriented manifold with orientation preserving $G$-action and $g : Y \to Z$ is equivariant, then $(g \circ f)_* = g_* \circ f_*$.

\item
If $\iota : X \to Y$ is an equivariant embedding, then $\iota^*(\iota_*(x)) = e^G(N) \cup x$ for all $x \in H^*_G(X; \Z)$, where $e^G(N) \in H^{\dim Y - \dim X}_G(X; \Z)$ is the $G$-equivariant Euler class of the normal bundle $N \to X$ of $\iota$.

\item
Let $X$, $X'$ and $Y$ be closed oriented dimensional manifolds with orientation preserving $G$-action, and $f : X \to Y$ and $f' : X' \to Y$ $G$-equivariant smooth map. Suppose $\dim X = \dim X'$. Then $(f \sqcup f')_* = f_* \oplus f'_*$.

\end{enumerate}

\subsection{Proof}
\label{sec:proof}

In this section, if the coefficient of a cohomology is $\Z$, then it will be omitted from the notation.

\subsubsection{Proof of Proposition \ref{prop:restriction_is_bijective}}

Proposition \ref{prop:restriction_is_bijective} follows from some lemmas. First of all, we relate the cohomology of $Y_{\mathrm{out}}$ with the equivariant cohomology of $T^2$.

\begin{lem} \label{lem:Y_out}
For any $k = 2, 3, 4, 6$ and $n \le 4$, the projection $D^6 \times T^2 \to T^2$ induces an isomorphism of groups
$$
H^n_{\Z_k}(T^2) \overset{\cong}{\longrightarrow} H^n_{\Z_k}(S^5 \times T^2)
\cong H^n((S^5 \times T^2)/\Z_k) = H^n(Y_{\mathrm{out}}).
$$
\end{lem}

\begin{proof}
We regard $S^5 \times T^2$ as the total space of the unit sphere bundle of the product $\Z_k$-equivariant complex vector bundle $\pi : \C^3 \times T^2 \to T^2$ associated to the representation of $\Z_k$ on $\C^3$. Then we apply the Gysin exact sequence
$$
\cdots \to
H^{n-6}_{\Z_k}(T^2) \to
H^n_{\Z_k}(T^2) \overset{\pi^*}{\to}
H^n_{\Z_k}(S^5 \times T^2) \to
H^{n-5}_{\Z_k}(T^2) \to
\cdots.
$$
Then the lemma follows from $H^m_{\Z_k}(T^2) = 0$ for $m < 0$. (The identification of the equivariant Chern class is possible, but is not necessary.)
\end{proof}

To understand $H^4_{\Z_k}(T^2)$, let $\widetilde{\Sing}_{T^2}$ be the set of points whose stabilizer groups are non-trivial
$$
\widetilde{\Sing}_{T^2}
= \{ p \in T^2 |\ \exists u \in \Z_k \backslash \{ 1 \}, up = p \}.
$$ 

\begin{lem} \label{lem:equiv_coh_T2}
For any $k = 2, 3, 4, 6$ and $n \ge 3$, the restriction induces an isomorphism of groups
$$
H^n_{\Z_k}(T^2) \overset{\cong}{\longrightarrow}
H^n_{\Z_k}(\widetilde{\Sing}_{T^2}).
$$
\end{lem}

\begin{proof}
For each $p \in \widetilde{\Sing}_{T^2}$, we choose a small closed disk $D^2(p) \subset T^2$ so that their (disjoint) union $D^2(\widetilde{\Sing}_{T^2}) = \bigsqcup_p D^2(p)$ is $\Z_k$-invariant. We remove the interior of $D^2(\widetilde{\Sing}_{T^2})$ to define ${}'T^2 = T^2 - \mathrm{Int}D^2(\tilde{\Sing}_{T^2})$. Then we can decompose the torus $T^2$ as the union of ${}'T^2$ and $D^2(\widetilde{\Sing}_{T^2})$, whose intersection is $\partial D^2(\widetilde{\Sing}_{T^2})$. We now look at the Mayer-Vietoris exact sequence
$$
\hspace{-0.3cm}
\cdot\cdot \to
H^n_{\Z_k}(T^2) \to
H^n_{\Z_k}({}'T^2) \oplus H^n_{\Z_k}(D^2(\widetilde{\Sing}_{T^2}))
\overset{\Delta}{\to}
H^n_{\Z_k}(\partial D^2(\widetilde{\Sing}_{T^2})) \to
H^{n+1}_{\Z_k}(T^2) \to 
\cdot\cdot,
$$
where $\Delta = i^* - j^*$ is the difference of the homomorphisms induced from the inclusions $i : \partial D^2(\widetilde{\Sing}_{T^2}) \to {}'T^2$ and $j : \partial D^2(\widetilde{\Sing}_{T^2}) \to D^2(\widetilde{\Sing}_{T^2})$. The intersection 
$$
{}'T^2 \cap D^2(\widetilde{\Sing}_{T^2})
= \partial D^2(\widetilde{\Sing}_{T^2})
= \bigsqcup_{p \in \widetilde{\Sing}_{T^2}} \partial D^2(p)
$$
is a union of circles. Notice that the action of $\Z_k$ on $\partial D^2(\widetilde{\Sing}_{T^2})$ is free, so that the quotient is again the disjoint union of a number of circles. It follows that 
$$
H^n_{\Z_k}(\partial D^2(\widetilde{\Sing}_{T^2})) \cong
H^n(\sqcup S^1) \cong 0
$$ 
for $n \ge 2$. Notice also that the action of $\Z_k$ on ${}'T^2$ is free. The quotient is a two-dimensional manifold with boundary. It follows that
$$
H^n_{\Z_k}({}'T^2) \cong 0
$$ 
for $n \ge 2$. In summary, for $n \ge 3$, the inclusion $D^2(\widetilde{\Sing}_{T^2}) \to T^2$ induces an isomorphism of groups
$$
H^n_{\Z_k}(T^2) \to H^n_{\Z_k}(\widetilde{\Sing}_{T^2}).
$$
Because $D^2(\widetilde{\Sing}_{T^2})$ is equivariantly homotopic to $\widetilde{\Sing}_{T^2}$, the lemma is proved.
\end{proof}

We next study the cohomology $H^n_{\Z_k}(\widetilde{X}') \cong H^n(X')$. Recall that
\begin{align*}
\widetilde{X} &= D^6 \times T^2, &
X &= (D^6 \times T^2)/\Z_k \\
\widetilde{X}' &= \widetilde{X} - \mathrm{Int}(D^8(\widetilde{\Sing})), &
X' &= \widetilde{X}'/\Z_k,
\end{align*}
in which $\widetilde{\Sing}$ consists of points with non-trivial stabilizer
$$
\widetilde{\Sing} = \{ x \in \widetilde{X} |\ \exists u \in \Z_k \backslash \{ 1 \}, ux = x \}
= \{ (0, 0, 0) \} \times \widetilde{\Sing}_{T^2},
$$
and $D^8(\widetilde{\Sing})$ is the $\Z_k$-invariant subspace given as the union of small closed disks around the points in $\widetilde{\Sing}$.

\begin{lem} \label{lem:X_prime_to_Y_in}
For $k = 2, 3, 4, 6$, the restriction induces the following isomorphisms
\begin{align*}
H^n_{\Z_k}(D^8(\widetilde{\Sing})) 
&\to H^n_{\Z_k}(\partial D^8(\widetilde{\Sing})),
& &(n \le 6) \\
H^n_{\Z_k}(\widetilde{X}) 
&\to H^n_{\Z_k}(\widetilde{X}'), 
& &(n \le 6) \\
H^n_{\Z_k}(\widetilde{X}) 
&\to H^n_{\Z_k}(D^8(\widetilde{\Sing})),
& &(n \ge 3) \\
H^n_{\Z_k}(\widetilde{X}') 
&\to H^n_{\Z_k}(\partial D^8(\widetilde{\Sing})) 
\cong H^n(Y_{\mathrm{in}}).
& & (3 \le n \le 6)
\end{align*}
\end{lem}

\begin{proof}
Note that $H^n_{\Z_k}(D^8(\widetilde{\Sing})) \cong H^n_{\Z_k}(\widetilde{\Sing})$ for any $n$, since $D^8(\widetilde{\Sing}) \simeq \widetilde{\Sing}$ equivariantly. Let $j : \partial D^8(\widetilde{\Sing}) \to D^8(\widetilde{\Sing})$ be the inclusion. The induced homomorphism
$$
H^n_{\Z_k}(\widetilde{\Sing})
\cong H^n_{\Z_k}(D^8(\widetilde{\Sing}))
\overset{j^*}{\to}
H^n_{\Z_k}(\partial D^8(\widetilde{\Sing}))
$$
agrees with that induced from the projection $\partial D^8(\widetilde{\Sing}) \to \widetilde{\Sing}$. We can think of $\partial D^8(\widetilde{\Sing})$ as the total space of the sphere bundle associated to a $\Z_k$-equivariant complex vector bundle $\widetilde{\Sing} \times \C^8 \to \widetilde{\Sing}$. In view of the Gysin exact sequence
$$
\cdots \to
H^{n-8}_{\Z_k}(\widetilde{\Sing}) \to
H^n_{\Z_k}(\widetilde{\Sing}) \to
H^n_{\Z_k}(\partial D^8(\widetilde{\Sing})) \to
H^{n-7}_{\Z_k}(\widetilde{\Sing}) \to
\cdots,
$$
we conclude that $j^* : H^n_{\Z_k}(D^8(\widetilde{\Sing})) \to H^n_{\Z_k}(\partial D^8(\widetilde{\Sing}))$ is an isomorphism for $n \le 6$.

\medskip

We can decompose $\widetilde{X}$ as the union of $\widetilde{X}'$ and $D^8(\widetilde{\Sing})$, whose intersection is $\partial D^8(\widetilde{\Sing})$. The Mayer-Vietoris sequence for this decomposition reads
$$
\cdot\cdot \to
H^n_{\Z_k}(\widetilde{X}) \overset{\imath^* \oplus \jmath^*}{\to}
H^n_{\Z_k}(\widetilde{X}') \oplus H^n_{\Z_k}(D^8(\widetilde{\Sing}))
\overset{\Delta_{\widetilde{X}}}{\to}
H^n_{\Z_k}(\partial D^8(\widetilde{\Sing})) \to
H^{n+1}_{\Z_k}(X) \to 
\cdot\cdot,
$$
where $\Delta_{\widetilde{X}} = i^* - j^*$ is the difference of the homomorphisms induced from the inclusions $i : \partial D^8(\widetilde{\Sing}) \to \widetilde{X}'$ and $j : \partial D^8(\widetilde{\Sing}) \to D^8(\widetilde{\Sing})$, and $\imath^*$ and $\jmath^*$ are induced from the inclusions $\imath : \widetilde{X}' \to \widetilde{X}$ and $\jmath : D^8(\widetilde{\Sing}) \to \widetilde{X}$. As is seen, $j^*$ is bijective for $n \le 6$, so that $\Delta_{\widetilde{X}}$ is surjective in degree $n \le 6$. Hence the Mayer-Vietoris exact sequence reduces to short exact sequences
$$
0 \to
H^n_{\Z_k}(\widetilde{X}) \overset{\imath^* \oplus \jmath^*}{\to}
H^n_{\Z_k}(\widetilde{X}') \oplus H^n_{\Z_k}(D^8(\widetilde{\Sing}))
\overset{\Delta_{\widetilde{X}}}{\to}
H^n_{\Z_k}(\partial D^8(\widetilde{\Sing})) \to
0
$$
for $n \le 6$. By doing a little homological algebra, we find that $\imath^* : H^n_{\Z_k}(\widetilde{X}) \to H^n_{\Z_k}(\widetilde{X}')$ is an isomorphism for $n \le 6$. (This can also be shown more directly by using the exact sequence for the pair $(\widetilde{X}, \widetilde{X}')$, the excision isomorphism and the Thom isomorphism.) We also find that, for $n \le 6$, the homomorphism $\jmath^* : H^n_{\Z_k}(\widetilde{X}) \to H^n_{\Z_k}(D^8(\widetilde{\Sing}))$ is injective (resp.\ surjective) if and only if $i^* : H^n_{\Z_k}(\widetilde{X}') \to H^n_{\Z_k}(\partial D^8(\widetilde{\Sing}))$ is injective (resp.\ surjective).

\medskip

To see the bijectivity of $\jmath^*$, let $\varpi : \widetilde{X} = D^6 \times T^2 \to T^2$ be the projection onto the second factor. This induces an equivariant homotopy equivalence $\widetilde{X} = D^6 \times T^2 \simeq T^2$. In the proof of Lemma \ref{lem:equiv_coh_T2}, it is shown that inclusion $\jmath_{T^2} : D^2(\widetilde{\Sing}_{T^2}) \to T^2$ induces an isomorphism
$$
\jmath_{T^2}^* : H^n_{\Z_k}(T^2) \to H^n_{\Z_k}(D^2(\widetilde{\Sing}_{T^2}))
$$
for $n \ge 3$. Without loss of generality, we can assume that $\varpi(D^8(\widetilde{\Sing})) = D^2(\widetilde{\Sing}_{T^2})$. Because $D^8(\widetilde{\Sing}) \simeq \widetilde{\Sing} \cong \widetilde{\Sing}_{T^2} \simeq D^2(\widetilde{\Sing}_{T^2})$ equivariantly, it is clear that 
$$
\varpi^* : H^n_{\Z_k}(D^2(\widetilde{\Sing}_{T^2})) \to
H^n_{\Z_k}(\varpi(D^8(\widetilde{\Sing})))
$$
is an isomorphism for all $n$. Now, for $n \ge 3$, we get a commutative diagram
$$
\begin{CD}
H^n_{\Z_k}(\widetilde{X}) @>{\jmath^*}>>
\overbrace{H^n_{\Z_k}(D^8(\widetilde{\Sing}))}^{
H^n_{\Z_k}(\widetilde{\Sing}_{T^2})
} 
\\
@A{\varpi^*}A{\cong}A @A{\cong}A{\varpi^*}A 
\\
H^n_{\Z_k}(T^2) 
@>{\jmath^*_{T^2}}>{\cong}>
\underbrace{H^n_{\Z_k}(D^2(\widetilde{\Sing}_{T^2}))}_{
H^n_{\Z_k}(\widetilde{\Sing}_{T^2})
}.
\end{CD}
$$
Therefore $\jmath^* : H^n_{\Z_k}(\widetilde{X}) \to H^n_{\Z_k}(D^8(\widetilde{\Sing}))$ is bijective for $n \ge 3$, and so is $i^* : H^n_{\Z_k}(\widetilde{X}') \to H^n_{\Z_k}(\partial D^8(\widetilde{\Sing}))$ for $3 \le n \le 6$.
\end{proof}

Recall $\partial \widetilde{X} = \partial (D^6 \times T^2) = S^5 \times T^2 \subset \widetilde{X}'$.

\begin{lem}
For $k = 2, 3, 4, 6$, the restriction induces the following isomorphism for $n \le 4$
$$
H^n_{\Z_k}(\widetilde{X}') \to H^n_{\Z_2}(\partial \widetilde{X})
 \cong H^n(Y_{\mathrm{out}}).
$$
\end{lem}

\begin{proof}
In view of the Gysin exact sequence in Lemma \ref{lem:Y_out}, the restriction induces an isomorphism 
$$
H^n_{\Z_k}(\widetilde{X}) \to H^n_{\Z_k}(\partial \widetilde{X})
$$
for $n \le 4$. Since $\partial \widetilde{X} \subset \widetilde{X}' \subset \widetilde{X}$, this isomorphism factors as follows
$$
H^n_{\Z_k}(\widetilde{X}) \to
H^n_{\Z_k}(\widetilde{X}') \to
H^n_{\Z_k}(\partial \widetilde{X}).
$$
The first homomorphism is bijective for $n \le 6$. Hence the second homomorphism is bijective for $n \le 4$.
\end{proof}

We summarize the homomorphisms so far in the following diagram:
$$
\begin{CD}
\overbrace{H^n_{\Z_2}(\partial \widetilde{X})}^{H^n(Y_{\mathrm{out}})}
@<{(n \le 4)}<< 
\overbrace{H^n_{\Z_k}(\widetilde{X}')}^{H^n(X')}
@>{(3 \le n \le 6)}>> 
\overbrace{H^n_{\Z_k}(\partial D^8(\widetilde{\Sing}))}^{
H^n(Y_{\mathrm{in}})}
\\
@A{n \le 4}AA 
@AA{n \le 6}A
@AA{n \le 6}A 
\\
H^n_{\Z_k}(T^2) @>>>
H^n_{\Z_k}(\widetilde{X}) @>>{(n \ge 3)}> 
H^n_{\Z_k}(D^8(\widetilde{\Sing})) 
\\
@V{n \ge 3}VV @. @|
\\
H^n_{\Z_k}(D^2(\widetilde{\Sing}_{T^2}))
@=
H^n_{\Z_k}(\widetilde{\Sing}_{T^2})
@=
H^n_{\Z_k}(\widetilde{\Sing}).
\end{CD}
$$
All the homomorphisms are induced from inclusions or projections. If an inequality such as $n \ge 3$ is put, then the homomorphism is bijective in degree $n \ge 3$. The homomorphism with inequality with parenthesis, such as $(n \ge 3)$, is bijective for $n \ge 3$, and this assumption is derived from those of other homomorphisms. The homomorphism without such an inequality is bijective for all $n \in \Z$.

\medskip

\subsubsection{Preliminary to proof of Proposition \ref{prop:maps}}
\label{subsec:preliminary}

To prove Proposition \ref{prop:maps}, we start with reducing the problem about $H^*(Y_{\mathrm{out}}) \cong H^*_{\Z_k}(S^5 \times T^2)$ to that about $H^*_{\Z_k}(T^2)$.

\medskip

Recall that $(S^5 \times T^2)/\Z_k$ is a fiber bundle over the lens space $S^5/\Z_k$ with fiber $T^2$. Now, letting $\Z_k$ acts on $\C^r$ in the same way for any $r$, we get a fiber bundle $(S^{2r-1} \times T^2)/\Z_k$ over $S^{2r-1}/\Z_k$ with fiber $T^2$. These fiber bundles form a direct system, and hence we get a map of $T^2$-bundles:
$$
\begin{CD}
(S^5 \times T^2)/\Z_k @>>>
\varinjlim_r (S^{2r-1} \times T^2)/\Z_k
@=
(E\Z_k \times T^2)/\Z_k \\
@VVV @VVV  @. \\
S^5/\Z_k @>>> \varinjlim_r S^{2r-1}/\Z_k
@= B\Z_k.
\end{CD}
$$
The $\Z_k$-action on $E\Z_k \times T^2$ above may be not exactly the same as that defining the homotopy quotient $E\Z_k \times_{\Z_k} T^2$. However, because the classifying space $E \Z_k \to B\Z_k$ is unique up to homotopy equivalence, the action of $u \in \Z_k$ on $\C$ given by the scalar multiplication $z \mapsto uz$ and that given by $z \mapsto \overline{u}z$ both lead to the classifying space $E\Z_k$. Thus, the cohomology of $(E\Z_k \times T^2)/\Z_k$ above is isomorphic to the Borel equivariant cohomology of $T^2$:
$$
H^n((E \Z_k \times T^2)/\Z_k) \cong H^n_{\Z_2}(T^2). \quad
(n \in \Z)
$$

\medskip

Now, because of the above map of fiber bundles, there are induced homomorphisms of the Leray-Serre spectral sequences and the corresponding filtrations of the cohomology of the total spaces. To be explicit, we have the filtration $F^{p, q}((S^5 \times T^2)/\Z_k)$ of the cohomology $H^{p+q}((S^5 \times T^2)/\Z_k)$
$$
H^{p+q}((S^5 \times T^2)/\Z_k)
= F^{0, p+q}((S^5 \times T^2)/\Z_k)
\supset 
F^{1, p+q-1}((S^5 \times T^2)/\Z_k) 
\supset \cdot\cdot.
$$
We also have the filtration $F^{p, q}_{\Z_k}(T^2)$ of $H^{p+q}(S^5 \times T^2)/\Z_k) \cong H^{p+q}_{\Z_k}(T^2)$
\begin{align*}
H^{p+q}_{\Z_k}(T^2)
= F^{0, p+q}_{\Z_k}(T^2)
\supset 
F^{1, p+q-1}_{\Z_k}(T^2)
\supset \cdots.
\end{align*}
Then the map $(S^5 \times T^2)/\Z_k \to (E\Z_k \times T^2)/\Z_k$ induces the homomorphism of the cohomology groups
$$
H^{p+q}_{\Z_k}(T^2) \to H^{p+q}((S^5 \times T^2)/\Z_k)
= H^{p+q}(Y_{\mathrm{out}})
$$
and the filtrations are preserved by this homomorphism. The Leray-Serre spectral sequence for $H^*_{\Z_k}(T^2)$ is computed in exactly the same way as in Sec.~\ref{subsec4.1}, and the graded quotients relevant to the fourth cohomology
\begin{align*}
&F^{2, 2}/F^{3, 1}, &
&F^{3, 1}/F^{4, 0}, &
&F^{4, 0}/F^{5, -1} = F^{4, 0}
\end{align*}
are isomorphic by the above homomorphism on the cohomology groups. As a result, to understand the homomorphisms
$$
F^4 \to H^4(Y_{\mathrm{out}}) \to F^2/F^3
$$ 
it is enough to understand 
$$
F^4 \to H^4_{\Z_k}(T^2) \to F^2/F^3.
$$

\bigskip

The above argument shows that $(S^5 \times T^2)/\Z_k \to (E\Z_k \times T^2)/\Z_k$ induces an isomorphism of fourth cohomology groups
$$
H^4_{\Z_k}(T^2) \to H^4((S^5 \times T^2)/\Z_k) = H^4(Y_{\mathrm{out}}).
$$
By Lemma \ref{lem:Y_out}, we also have an isomorphism
$$
H^4_{\Z_k}(T^2) \to H^4_{\Z_k}(S^5 \times T^2)
\cong H^4((S^5 \times T^2)/\Z_k)
$$
induced from the projection $S^5 \times T^2 \to T^2$. These isomorphisms are the same, as a result of the following lemma.

\begin{lem}
We define homomorphisms 
\begin{align*}
\iota &: (S^5 \times T^2)/\Z_k \qquad\quad\to (E\Z_k \times T^2)/\Z_k, &
\iota([x, t]) &= [i(x), t], \\
\varpi &: (E\Z_k \times S^5 \times T^2)/\Z_k \to (S^5 \times T^2)/\Z_k, &
\varpi([\xi, x, t]) &= [x, t], \\
\pi &: (E\Z_k \times S^5 \times T^2)/\Z_k \to (E\Z_k \times T^2)/\Z_k, &
\pi([\xi, x, t]) &= [\xi, t],
\end{align*}
where $i : S^5 \to S^\infty$ is the natural inclusion. Then $(\iota \circ \varpi)^* = \pi^*$ on the cohomology groups.
\end{lem}

\begin{proof}
We have the diagram
$$
\begin{CD}
(S^5 \times T^2)/\Z_k @>{\iota}>>
\varinjlim_r (S^{2r-1} \times T^2)/\Z_k \\
@A{\mathrm{proj}}A{\varpi}A @| \\
(E\Z_k \times S^5 \times T^2)/\Z_k @>{\mathrm{proj}}>{\pi}>
(E\Z_k \times T^2)/\Z_k.
\end{CD}
$$
This is not commutative. But, $\varpi$ induces the isomorphism $H^*((S^5 \times T^2)/\Z_k) \cong H^*_{\Z_k}(S^5 \times T^2)$. This is because $\varpi : (E\Z_k \times S^5 \times T^2)/\Z_k \to (S^5 \times T^2)/\Z_k$ is a fiber bundle with contractible fiber $E\Z_k$, and so $\varpi$ is a homotopy equivalence. Let us define a section of $\varpi$ as follows
\begin{align*}
\sigma &: (S^5 \times T^2)/\Z_k \to 
(E\Z_k \times S^5 \times T^2)/\Z_k, &
\sigma([x, t]) := [i(x), x, t].
\end{align*}
Since $\varpi \circ \sigma = 1$ and $\varpi^*$ is an isomorphism, the induced homomorphism $\sigma^*$ is also an isomorphism inverse to $\varpi^*$. It is clear that $\pi \circ \sigma = \iota$. From this, we now have $\iota^* = \sigma^* \circ \pi^* = (\varpi^*)^{-1}$, and hence $\varpi^* \circ \iota^* = \pi^*$.
\end{proof}

\medskip

Recall finally that the inclusion $\widetilde{\Sing}_{T_2} \to T^2$ induces
$$
H^n_{\Z_k}(\widetilde{\Sing}_{T^2}) \to
H^n_{\Z_k}(T^2),
$$
which is isomorphic for $n \ge 3$ by Lemma \ref{lem:equiv_coh_T2}. The $\Z_k$-space $\widetilde{\Sing}_{T^2}$ is a union of the $\Z_k$-spaces of the form $\Z_k/\Z_\ell$, where $\Z_\ell \subset \Z_k$ is a subgroup: 
$$
\widetilde{\Sing}_{T^2}
= 
\left\{
\begin{array}{ll}
\Z_2/\Z_2 \sqcup \Z_2/\Z_2 \sqcup \Z_2/\Z_2 \sqcup \Z_2/\Z_2, & (k = 2) \\
\Z_3/\Z_3 \sqcup \Z_3/\Z_3 \sqcup \Z_3/\Z_3, & (k = 3) \\
\Z_4/\Z_4 \sqcup \Z_4/\Z_4 \sqcup \Z_4/\Z_2, & (k = 4) \\
\Z_6/\Z_6 \sqcup \Z_6/\Z_3 \sqcup \Z_6/\Z_2. & (k = 6)
\end{array}
\right.
$$
From this and the isomorphism $H^4_{\Z_k}(\Z_k/\Z_\ell) \cong H^4_{\Z_\ell}(\pt)$, we get the same description of $H^4(Y_{\mathrm{out}}) \cong H^4(X') \cong H^4(Y_{\mathrm{in}})$ as given in \eqref{F}:
\begin{align*}
H^4(Y_{\mathrm{out}})
&\cong 
H^4_{\Z_k}(T^2)
\cong
H^4_{\Z_k}(\widetilde{\Sing}_{T^2})
\cong
\left\{
\begin{array}{ll}
\Z/2 \oplus \Z/2 \oplus \Z/2 \oplus \Z/2, & (k = 2) \\
\Z/3 \oplus \Z/3 \oplus \Z/3, & (k = 3) \\
\Z/4 \oplus \Z/4 \oplus \Z/2, & (k = 4) \\
\Z/6 \oplus \Z/3 \oplus \Z/2. & (k = 6)
\end{array}
\right.
\end{align*}
Let $\bar{n} \in \Z/\ell$ be the image of $n \in \Z$ under the quotient $\Z \to \Z/\ell$. In the description above, the `diagonal embedding' $\Z/k \to \bigoplus \Z/\ell$ is defined as the homomorphism that carries $\bar{1} \in \Z/k$ to $(\bar{1}) \in \bigoplus \Z/\ell$. The `sum' $\bigoplus \Z/\ell \to \Z/k$ is defined as the homomorphism that carries $(\bar{n}) \in \bigoplus \Z/\ell$ to $\sum \frac{k}{\ell} \bar{n} \in \Z/k$.

\subsubsection{Proof of Proposition \ref{prop:maps} (a)}

In view of Subsection \ref{subsec:preliminary}, to prove Proposition \ref{prop:maps} (a), we study the homomorphism
$$
F^4 \to H^4_{\Z_k}(T^2) \cong H^4_{\Z_k}(\widetilde{\Sing}_{T^2}),
$$
where the filtration $F^p = F^{p, 4-p}_{\Z_k}(T^2)$ is that for the Leray-Serre spectral sequence computing $H^4_{\Z_k}(T^2)$. Its $E_2$-term is $E_2^{p, q} = H^p(B\Z_k; \underline{H^q(T^2)})$, and 
\begin{align*}
F^4 
&= E^{4, 0}_\infty = E^{4, 0}_2 = H^4(B\Z_k; H^0(T^2)) 
= H^4_{\Z_k}(\pt) \cong \Z/k.
\end{align*}

The following is the counterpart of Proposition \ref{prop:maps} (a):

\begin{prop}
The homomorphism
$$
H^4_{\Z_k}(\pt) = F^4 \to H^4_{\Z_k}(T^2) \overset{\cong}{\to}
H^4_{\Z_k}(\widetilde{\Sing}_{T^2})
$$
is induced from $\widetilde{\Sing}_{T^2} \to \pt$, and agrees with the `diagonal embedding'.
\end{prop}

\begin{proof}
Let $\pi : T^2 \to \pt$ be the unique ($\Z_k$-equivariant) map. This induces a homomorphism of the Leray-Serre spectra sequences computing the equivariant cohomology of $T^2$ and $\pt$. Then we have the commutative diagram
$$
\begin{CD}
F^{4, 0}_{\Z_k}(T^2) @>>> H^4_{\Z_k}(T^2) \\
@A{\pi^*}A{\cong}A @AA{\pi^*}A \\
F^{4, 0}_{\Z_k}(\pt) @= H^4_{\Z_k}(\pt).
\end{CD}
$$
As a result, the homomorphism $F^4 \to H^4_{\Z_k}(T^2)$ of our interests is the same as $\pi^* : H^4_{\Z_k}(\pt) \to H^4_{\Z_k}(T^2)$. The composition
$$
H^4_{\Z_k}(\pt) \overset{\pi^*}{\to} 
H^4_{\Z_k}(T^2) \to
H^4_{\Z_k}(\widetilde{\Sing}_{T^2})
$$
is clearly induced from the unique map $\widetilde{\Sing}_{T^2} \to \pt$, which induces the ring homomorphism $H^*_{\Z_k}(\pt) \to H^*_{\Z_k}(\widetilde{\Sing}_{T^2})$, carrying the ring unit to the ring unit. Since the ring unit in $H^*_{\Z_k}(\widetilde{\Sing}_{T^2}) = \oplus H^*_{\Z_\ell}(\pt)$ is the direct sum of the ring units of $H^*_{\Z_\ell}(\pt)$, the basis $\bar{1} \in H^4_{\Z_k}(\pt) \cong \Z/k$ is mapped to $(\bar{1}) \in H^4_{\Z_k}(\widetilde{\Sing}_{T^2}) \cong \bigoplus \Z/\ell$, and the homomorphism in question is the `diagonal embedding'.
\end{proof}

\subsubsection{Proof of Proposition \ref{prop:maps} (b)}

In view of Subsection \ref{subsec:preliminary}, to prove Proposition \ref{prop:maps} (b), it suffices to see the homomorphism
$$
H^4_{\Z_k}(\widetilde{\Sing}_{T^2}) \cong
H^4_{\Z_k}(T^2) \to F^2/F^3,
$$
where the filtration $F^p = F^{p, 4-p}_{\Z_k}(T^2)$ is that for the Leray-Serre spectral sequence computing $H^4_{\Z_k}(T^2)$, whose $E_2$-term is $E_2^{p, q} = H^p(B\Z_k; \underline{H^q(T^2)})$, and 
\begin{align*}
F^2/F^3
&= E^{2, 2}_\infty = E^{2, 2}_2 = H^2(B\Z_k; H^2(T^2)) 
= H^2_{\Z_k}(\pt) \cong \Z/k.
\end{align*}
Then the counterpart of Proposition \ref{prop:maps} (b) will be shown by identifying the homomorphism in question with the push-forward along $\pi : T^2 \to \pt$. Thus, we start with this identification.

\medskip

For the identification, we summarize a property of the $\Z_k$-equivariant cohomology of $T^2$, which will also be used in the proof of the counterpart of Proposition \ref{prop:maps} (b).

\begin{lem} \label{lem:higher_degree}
For $k = 2, 3, 4, 6$, the following holds true.
\begin{itemize}
\item[(a)]
$H^n_{\Z_k}(T^2) = 0$ if $n \ge 3$ is odd.

\item[(b)]
Suppose that $n \in \Z$ is even. The homomorphism 
$$
t : \ H^n_{\Z_k}(T^2) \longrightarrow H^{n+2}_{\Z_k}(T^2)
$$
multiplying the basis $t \in H^2_{\Z_k}(\pt)$ is surjective for $n = 2$, and is bijective for $n \ge 4$. In addition, the following diagram is commutative
$$
\begin{CD}
H^2_{\Z_k}(T^2) @>{t}>{\mathrm{surj}}>
H^4_{\Z_k}(T^2) @>{t}>{\cong}>
H^6_{\Z_k}(T^2) @>{t}>{\cong}> \cdots \\
@VVV @VV{\cong}V @VV{\cong}V @. \\
H^2_{\Z_k}(\widetilde{\Sing}_{T^2}) @>{t}>{\cong}>
H^4_{\Z_k}(\widetilde{\Sing}_{T^2}) @>{t}>{\cong}>
H^6_{\Z_k}(\widetilde{\Sing}_{T^2}) @>{t}>{\cong}> \cdots,
\end{CD}
$$
where the vertical maps are induced from the inclusion $\widetilde{\Sing}_{T^2} \subset T^2$.

\end{itemize}
\end{lem}

\begin{proof}
The lemma follows from the Mayer-Vietoris sequence in Lemma \ref{lem:equiv_coh_T2}: The vanishing in (a) is easy to see. For (b), we find that the homomorphism $H^2_{\Z_k}(T^2) \to  H^2_{\Z_k}(\widetilde{\Sing}_{T^2})$ is surjective. Consequently, we get a commutative diagram
$$
\begin{CD}
H^2_{\Z_k}(T^2) @>{t}>>
H^4_{\Z_k}(T^2) @>{t}>>
H^6_{\Z_k}(T^2) @>{t}>> \cdots \\
@V{\mathrm{surj}}VV @VV{\cong}V @VV{\cong}V @. \\
H^2_{\Z_k}(\widetilde{\Sing}_{T^2}) @>{t}>>
H^4_{\Z_k}(\widetilde{\Sing}_{T^2}) @>{t}>>
H^6_{\Z_k}(\widetilde{\Sing}_{T^2}) @>{t}>> \cdots.
\end{CD}
$$
It is easy to see that $t : H^n_{\Z_k}(\widetilde{\Sing}_{T^2}) \to H^{n+2}(\widetilde{\Sing}_{T^2})$ is an isomorphism for $n \ge 1$. 
\end{proof}

In general, if a compact Lie group $G$ acts on a closed oriented $d$-dimensional manifold $X$ preserving its orientation, then we can define a homomorphism (called the edge homomorphism)
$$
\epsilon : H^n_{G}(X) \to H^{n-d}_{G}(X)
$$
for $n \in \Z$ as the composition of 
$$
H^n_{G}(X) \to F^{n-d,d}/F^{n-d+1, d-1}
= E^{n-d, d}_\infty \subset E^{n-d, d}_2 = H^{n-d}_G(\pt),
$$
where $F^{p, q}$ are the filtration in the Leray-Serre spectral sequence. It seems that the edge homomorphism is believed to be identical to the push-forward $\pi_* : H^n_G(X) \to H^{n-d}(\pt)$, while its proof cannot be found in the literature at hand. However, at least in our case, we can provide a proof of the coincidence.

\begin{lem}
For $k = 2, 3, 4, 6$ and $n \in \Z$, the edge homomorphism agrees with the push-forward along $\pi : T^2 \to \pt$,
$$
\epsilon = \pi_* : \ H^n_{\Z_k}(T^2) \to H^{n-2}_{\Z_k}(\pt).
$$
\end{lem}

\begin{proof}
In the case that $n < 2$, we have $H^{n-2}_{\Z_k}(\pt) = 0$, so that the coincidence of $\epsilon = 0$ and $\pi_* = 0$ is clear. In the case that $n \ge 3$ is odd, we also have $\epsilon = 0 = \pi_*$ by Lemma \ref{lem:higher_degree} (a). In the case that $n > 2$ is even, then $t^{\frac{n-2}{2}} : H^2_{\Z_k}(T^2) \to H^n_{\Z_k}(T^2)$ is surjective by Lemma \ref{lem:higher_degree} (b). Recall the projection formula of the push-forward $\pi_*(\pi^*a \cup x) = a \cup \pi_*(x)$ for $a \in H^*_{\Z_k}(\pt)$ and $x \in H^*_{\Z_k}(T^2)$. Since the Leray-Serre spectral sequence is compatible with the $H^*_{\Z_k}(\pt)$-module structure on the equivariant cohomology, the edge homomorphism also satisfies the projection formula $\epsilon(\pi^*a \cup x) = a \cup \epsilon(x)$. As a result, the case that $n > 2$ even follows from that of $n = 2$. Hence the present lemma will be completed by showing $\epsilon = \pi_*$ in the case that $n = 2$.

Because $H^0_{\Z_k}(\pt) = \Z$, both $\epsilon$ and $\pi_*$ are trivial on the torsion subgroups of $H^2_{\Z_k}(T^2)$. Hence it suffices to see $\epsilon = \pi_*$ on the free part. The free part injects into the cohomology $H^2_{\Z_k}(T^2; \R)$ with real coefficients. The homomorphism induced from the inclusion $\Z \to \R$ is a cohomology transformation preserving the Thom isomorphisms. It follows that the push-forward in the integral cohomology is compatible with that in the real cohomology, so that the problem is reduced to showing the coincidence in the equivariant cohomology with real coefficients. 

By the help of the Leray-Serre spectral sequence, we can show that the homomorphism $f : H^2_{\Z_k}(T^2; \R) \to H^2(T^2; \R)$ forgetting the group action induces an isomorphism in the present case. Similarly, $f : H^0_{\Z_k}(\pt; \R) \to H^0(\pt; \R)$ is an isomorphism. The homomorphism $f$ is a cohomology transformation which preserves the Thom isomorphisms. Hence the push-forward in equivariant real cohomology is compatible with that in the real cohomology. In summary, it suffices to show that
$$
\epsilon = \pi_* : \underbrace{H^2(T^2; \R)}_{\R} \to 
\underbrace{H^0(\pt; \R)}_{\R}.
$$
In view of the Leray-Serre spectral sequence for $T^2 \to \pt$, the edge homomorphism is nothing but the isomorphism $H^2(T^2; \R) \cong \R$ induced from the orientation on $T^2$. In general, the Thom isomorphism in real cohomology is inverse to the fiber integration. Hence the push-forward is also an isomorphism $H^2(T^2; \R) \cong \R$ induced from the orientation, and we conclude $\epsilon = \pi_*$.
\end{proof}

Now, our task is to study the push-forward $\pi_* : H^4_{\Z_k}(T^2) \to H^2_{\Z_k}(\pt)$.

\begin{lem} \label{lem:key_diagram}
The following diagram is commutative:
$$
\begin{CD}
H^4_{\Z_k}(T^2) @>{\pi_*}>> H^2_{\Z_k}(\pt) \\
@V{j^*}V{\cong}V @V{\cong}V{t}V \\
H^4_{\Z_k}(\widetilde{\Sing}_{T^2}) @>>{\varpi_*}> H^4_{\Z_k}(\pt),
\end{CD}
$$
where $\pi : T^2 \to \pt$ and $\widetilde{\Sing}_{T^2} \to \pt$ are the maps to the point, $j : \widetilde{\Sing}_{T^2} \to T^2$ is the inclusion, and $t$ is the multiplication with the basis $t \in H^2_{\Z_k}(\pt)$.
\end{lem}

\begin{proof}
We consider the following diagram:
$$
\xymatrix{
H^4_{\Z_k}(T^2) \ar[rr]^{\pi_*} \ar[dd]_{j^*}^{\cong} & & 
H^2_{\Z_k}(\pt) \ar[dd]^{t}_{\cong} \\
& H^2_{\Z_k}(\widetilde{\Sing}_{T^2}) 
\ar[lu]_{j_*} \ar[ru]^{\varpi_*} \ar[ld]^{t}_{\cong} & \\
H^4_{\Z_k}(\widetilde{\Sing}_{T^2}) \ar[rr]_{\varpi_*} & &
H^4_{\Z_k}(\pt)
}
$$
By a property mentioned in Section \ref{sec:equivariant_coh}, the push-forward along the embedding $j : \widetilde{\Sing}_{T^2} \to T^2$ satisfies
$$
j^*(j_*(x)) = e^{\Z_k}(N)x
$$
for $x \in H^2_{\Z_k}(\widetilde{\Sing}_{T^2})$, where $e^{\Z_k}(N) \in H^2_{\Z_k}(\widetilde{\Sing}_{T^2})$ is the equivariant Euler class of the normal bundle $N \to \widetilde{\Sing}_{T^2}$ of $j$. In the present setup, the tangent bundle of $T^2$ is isomorphic to the product $\Z_k$-equivariant complex line bundle associated to the representation of $\Z_k \subset U(1)$. It follows that the Euler class is the (pull-back of the) generator $t \in H^2_{\Z_k}(\pt)$. This induces an isomorphism $t : H^2_{\Z_k}(\widetilde{\Sing}_{T^2}) \to H^4_{\Z_k}(\widetilde{\Sing}_{T^2})$. The pull-back $j^* : H^4_{\Z_k}(T^2) \to H^4_{\Z_k}(\widetilde{\Sing}_{T^2})$ is bijective by Lemma \ref{lem:equiv_coh_T2}. Hence the push-forward $j_* : H^2_{\Z_k}(\widetilde{\Sing}_{T^2}) \to H^4_{\Z_k}(T^2)$ turns out to be bijective. Noting that $\pi \circ j = \varpi$ and the push-forward is a homomorphism of modules over the cohomology ring $H^*_{\Z_k}(\pt)$, we see the commutativity of the diagram.
\end{proof}

\begin{lem} \label{lem:push_forward_orbit}
Let $\Z_k$ be a cyclic group of order $k$, and $\Z_\ell \subset \Z_k$ a subgroup. We regard $\Z_k/\Z_\ell$ as a $\Z_k$ space by the multiplication in $\Z_k$. Then the homomorphism given by the push-forward along $\Z_k/\Z_\ell \to \pt$ 
$$
\underbrace{H^0_{\Z_\ell}(\pt)}_{\Z} \cong
H^0_{\Z_k}(\Z_k/\Z_\ell) \to
\underbrace{H^0_{\Z_k}(\pt)}_{\Z}
$$
is the multiplication with $k/\ell$.
\end{lem}

\begin{proof}
Let $\varpi : \Z_k/\Z_\ell \to \pt$ be the map to the point. The push-forward in equivariant cohomology is compatible with that in the ordinary cohomology through the homomorphism forgetting the group action. This means that we have the commutative diagram
$$
\begin{CD}
\overbrace{H^0_{\Z_k}(\Z_k/\Z_\ell)}^{\Z} @>{\varpi_*}>> 
\overbrace{H^0_{\Z_k}(\pt)}^{\Z} \\
@V{f}VV @VV{f}V \\
H^0(\Z_k/\Z_\ell) @>{\varpi_*}>> \underbrace{H^0(\pt)}_{\Z}.
\end{CD}
$$
The homomorphism $f : H^0_{\Z_k}(\pt) \to H^0(\pt)$ is the identity map $\Z \to \Z$. Since $\Z_k/\Z_\ell$ consists of $k/\ell$ points, we have $H^0(\Z_k/\Z_\ell) \cong \Z^{\oplus k/\ell}$, and the homomorphism $f : H^0_{\Z_k}(\Z_k/\Z_\ell) \to H^0(\Z_k/\Z_\ell)$ is easily identified with the diagonal embedding $\Z \to \Z^{\oplus k/\ell}$. It is also easy to see that $\pi_* : H^0(\Z_k/\Z_\ell) \to H^0(\pt)$ is the sum $(n_1, \ldots, n_{k/\ell}) \mapsto n_1 + \cdots + n_{k/\ell}$. Hence $\varpi_*$ in question is the homomorphism $\Z \to \Z$ multiplying $k/\ell$.
\end{proof}

We are now in the position to prove the counterpart of Proposition \ref{prop:maps} (b).

\begin{prop}
For $k = 2, 3, 4, 6$, the composition of the homomorphisms
$$
H^4_{\Z_k}(\widetilde{\Sing}_{T^2}) 
\cong
H^4_{\Z_k}(T^2) \overset{\epsilon}{\to} H^4_{\Z_k}(\pt) = \Z/k
$$
is the `sum' homomorphism.
\end{prop}

\begin{proof}
If $n \ge 2$ is even, then we have 
$$
H^n_{\Z_k}(\widetilde{\Sing}_{T^2})
\cong
\left\{
\begin{array}{ll}
\Z/2 \oplus \Z/2 \oplus \Z/2 \oplus \Z/2, & (k = 2) \\
\Z/3 \oplus \Z/3 \oplus \Z/3, & (k = 3) \\
\Z/4 \oplus \Z/4 \oplus \Z/2, & (k = 4) \\
\Z/6 \oplus \Z/3 \oplus \Z/2, & (k = 6)
\end{array}
\right.
$$
and these isomorphisms are preserved by the isomorphism $t : H^n_{\Z_k}(\widetilde{\Sing}_{T^2}) \to H^{n+2}_{\Z_k}(\widetilde{\Sing}_{T^2})$. As is seen, the edge homomorphism $\epsilon$ agrees with the push-forward along $\pi : T^2 \to \pt$. Then, by Lemma \ref{lem:key_diagram}, it suffices to describe the push-forward $\varpi_* : H^4_{\Z_k}(\widetilde{\Sing}_{T^2}) \to H^4_{\Z_k}(\pt)$ along $\varpi : \widetilde{\Sing}_{T^2} \to \pt$. In view of the commutative diagram
$$
\begin{CD}
H^0_{\Z_k}(\widetilde{\Sing}_{T^2})
@>{t}>{\mathrm{surj}}>
H^2_{\Z_k}(\widetilde{\Sing}_{T^2})
@>{t}>{\cong}>
H^4_{\Z_k}(\widetilde{\Sing}_{T^2}) \\
@V{\varpi_*}VV @V{\varpi_*}VV @V{\varpi_*}VV \\
\underbrace{H^0_{\Z_k}(\pt)}_{\Z}
@>{t}>{\mathrm{surj}}>
\underbrace{H^2_{\Z_k}(\pt)}_{\Z/k}
@>{t}>{\cong}>
\underbrace{H^4_{\Z_k}(\pt)}_{\Z/k},
\end{CD}
$$
and the description of the push-forward $H^0_{\Z_k}(\Z_k/\Z_\ell) \to H^0_{\Z_k}(\pt)$ in Lemma \ref{lem:push_forward_orbit}, we find that $\varpi_* : H^4_{\Z_k}(\widetilde{\Sing}_{T^2}) \to H^4_{\Z_k}(\pt)$ is the `sum' homomorphism.
\end{proof}

\bibliographystyle{ytphys}
\baselineskip=.85\baselineskip
\let\bbb\bibitem\def\bibitem{\itemsep1pt\bbb}
\bibliography{3plane-refs}

\end{document}